%% file: HIG-20-001_temp.tex
\pdfoutput=1
\documentclass[11pt,twoside,a4paper,cmspaper,final,collab]{cms-tdr}
\def\svnVersion{41b7c0b}\def\svnDate{2024/05/30}\def\cmsCernNoTag{CERN-EP-2023-270}\def\cmsCernDate{\today}\def\cmsMessage{Published in Physical Review D as \href{http://dx.doi.org/10.1103/PhysRevD.109.092011}{\doi{10.1103/PhysRevD.109.092011}.}}
\begin{document}\cmsNoteHeader{HIG-20-001}

\ifthenelse{\boolean{cms@external}}{\providecommand{\cmsTable}[1]{#1}}{\providecommand{\cmsTable}[1]{\resizebox{\textwidth}{!}{#1}}}
\newlength\cmsTabSkip\setlength{\cmsTabSkip}{0.5ex}
\newcommand{\Znn}{\ensuremath{\PZ(\PGn\PGn)}}
\newcommand{\ZnnH}{\ensuremath{\PZ(\PGn\PGn)\PH}}
\newcommand{\ZllH}{\ensuremath{\PZ(\Pell\Pell)\PH}}
\newcommand{\ZmmH}{\ensuremath{\PZ(\PGm\PGm)\PH}}
\newcommand{\ZeeH}{\ensuremath{\PZ(\Pe\Pe)\PH}}
\newcommand{\WmnH}{\ensuremath{\PW(\PGm\PGn)\PH}}
\newcommand{\WenH}{\ensuremath{\PW(\Pe\PGn)\PH}}
\newcommand{\WtoLN}{\ensuremath{\PW\to\Pell\PGn}}
\newcommand{\ZtoNN}{\ensuremath{\PZ\to\PGn\PAGn}}
\newcommand{\ZtoLL}{\ensuremath{\PZ\to\Pell\Pell}}
\newcommand{\jj}{\ensuremath{\text{jj}}}
\newcommand{\Mjj}{\ensuremath{M(\jj)}}
\newcommand{\ptjj}{\ensuremath{{\pt}(\jj)}}
\newcommand{\dRJJ}{\ensuremath{\Delta R(\jj)}}
\newcommand{\dphiZH}{\ensuremath{\Delta\phi(\PZ,\PH)}}
\newcommand{\dphiZjj}{\ensuremath{\Delta\phi(\PZ,\jj)}}
\newcommand{\dphiWjj}{\ensuremath{\Delta\phi(\PW,\jj)}}
\newcommand{\dphiVjj}{\ensuremath{\Delta\phi(\PV,\jj)}}
\newcommand{\dphiMJ}{\ensuremath{\Delta\phi(\text{\ptvecmiss, j})}}
\newcommand{\dphiMtkM}{\ensuremath{\Delta\phi(\ptvecmiss,\ptvecmiss{}_{\text{trk}})}}
\newcommand{\ptV}{\ensuremath{\pt(\PV)}}
\newcommand{\ptH}{\ensuremath{\pt(\PH)}}
\newcommand{\ptM}{\ensuremath{\pt^{\text{miss}}}}
\newcommand{\ptMtrk}{\ensuremath{p_{\text{T,trk}}^{\text{miss}}}}
\newcommand{\Naj}{\ensuremath{N_{\text{aj}}}}
\newcommand{\etaTF}{\ensuremath{\abs{\eta}  < 2.5}}
\newcommand{\minMETMHT}{\ensuremath{\min(\ptM, \mht)}}
\newcommand{\mH}{\ensuremath{m_\PH}}
\newcommand{\twol}{\text{2-lepton}}
\newcommand{\onel}{\text{1-lepton}}
\newcommand{\zerol}{\text{0-lepton}}
\newcommand{\DeepAKviii}{\textsc{DeepAK8}\xspace}
\newcommand{\DeepCSV}{\textsc{DeepCSV}\xspace}

\ifthenelse{\boolean{cms@external}}{\providecommand{\cmsLeft}{upper\xspace}}{\providecommand{\cmsLeft}{left\xspace}}
\ifthenelse{\boolean{cms@external}}{\providecommand{\cmsRight}{lower\xspace}}{\providecommand{\cmsRight}{right\xspace}}

\cmsNoteHeader{HIG-20-001} 
\title{Measurement of simplified template cross sections of the Higgs boson produced in association with \texorpdfstring{\PW or \PZ}{W or Z} bosons in the  \texorpdfstring{$\PH\to \PQb\PAQb$}{H to \PQb bbar} decay channel in proton-proton collisions at \texorpdfstring{$\sqrt{s}=13\TeV$}{sqrt(s) = 13 TeV}}

\author*[inst1]{Christophorus Grab}
\date{\today}

\abstract{Differential cross sections are measured for the standard model Higgs boson produced in association with vector bosons (\PW, \PZ) and decaying to a pair of \PQb quarks. Measurements are performed within the framework of the simplified template cross sections. The analysis relies on the leptonic decays of the \PW and \PZ bosons, resulting in final states with 0, 1, or 2 electrons or muons. The Higgs boson candidates are either reconstructed from pairs of resolved \PQb-tagged jets, or from single large-radius jets containing the particles arising from two \PQb quarks. Proton-proton collision data at $\sqrt{s}=13\TeV$, collected by the CMS experiment in 2016--2018 and corresponding to a total integrated luminosity of 138\fbinv, are analyzed. The inclusive signal strength, defined as the product of the observed production cross section and branching fraction relative to the standard model expectation, combining all analysis categories, is found to be $\mu=1.15^{+0.22}_{-0.20}$. This corresponds to an observed (expected) significance of 6.3 (5.6) standard deviations.}

\hypersetup{%
pdfauthor={CMS Collaboration},%
pdftitle={Measurement of simplified template cross sections of the Higgs boson produced in association with W or Z bosons in the H to bb-bar decay channel in proton-proton collisions at sqrt(s)=13 TeV},%
pdfsubject={CMS},%
pdfkeywords={CMS, Higgs, VHbb}}
\maketitle 

\section{Introduction}
The discovery of a Higgs boson (\PH) with a mass near 125\GeV by the ATLAS~\cite{Aad:2012tfa} and CMS~\cite{Chatrchyan:2012ufa,Chatrchyan:2013lba} Collaborations at the CERN LHC was a major milestone in the understanding of electroweak (EW) symmetry breaking in the standard model (SM) of particle physics~\cite{Englert:1964et,Higgs:1964ia,Higgs:1964pj,Guralnik:1964eu,Higgs:1966ev,Kibble:1967sv}. Since its observation, most of the Higgs boson production modes and many decay channels predicted by the SM have been measured. In particular, decays of the Higgs boson into the $\PGg\PGg$, $\PZ\PZ$, $\PW\PW$, $\PGt\PGt$, and $\PQb\PAQb$ channels have been observed~\cite{Aad:2014eha,Khachatryan:1728107,Aad:2014eva,Chatrchyan:2013mxa,ATLAS:2014aga,Aad:2015ona,Chatrchyan:1633401,Aad:2015vsa,Chatrchyan:1643937,Sirunyan:2273798,Sirunyan:2018egh,Aaboud:2018zhk}. Higgs boson production in association with a top quark-antiquark pair ($\ttbar$)~\cite{Sirunyan:2018hoz,Aaboud:2018urx} has also been observed. To date, all the measured properties~\cite{HIGG-2021-23,CMS-HIG-22-001} are compatible with the hypothesis that this particle is the Higgs boson predicted by the SM. Evidence for the Higgs boson decaying to a pair of muons has been reported by the CMS Collaboration, which further supports compatibility with the SM hypothesis~\cite{Khachatryan:200904363}.

For an SM Higgs boson with a mass ($\mH$) of 125\GeV, the largest Yukawa coupling that is directly accessible by studying a specific Higgs boson decay channel is the coupling between the Higgs boson and \PQb quarks. The Yukawa couplings to fermions are proportional to the fermion masses. The Higgs boson decay into pairs of \PQb quarks is kinematically allowed, and out of all Higgs boson decay channels, the $\PH\to\PQb\PAQb$ decay has the largest branching fraction. This decay was previously observed by the ATLAS and CMS Collaborations~\cite{Aaboud:2018zhk,Sirunyan:2018kst}. Using the data collected at $\sqrt{s}=7$, 8, and 13\TeV in 2011--2017, the CMS measurement of the $\PH\to\PQb\PAQb$ decay in the vector boson (\PV, $\PV=\PW, \PZ$) associated production mode ($\PV\PH$) had a significance of 4.8 standard deviations over the background-only hypothesis. The corresponding measured signal strength, defined as the production cross section times branching fraction relative to its SM expectation, was $\mu=1.01 \pm 0.22$. In combination with other production modes, the measured signal strength in the $\PH\to\PQb\PAQb$ channel was $1.04 \pm 0.20$, corresponding to an observed 
significance of 5.6 standard deviations~\cite{Sirunyan:2018kst}.

The large size of the data set delivered by the LHC between 2016 and 2018  permits both the ATLAS and CMS Collaborations to make detailed measurements of the kinematic properties of VH production using the $\PH\to\PQb\PAQb$ decay channel. The analysis targets the following decay channels of the \PW or \PZ boson: $\ZtoNN$, $\WtoLN$, and $\ZtoLL$. These decay modes are referred to as $\zerol$, $\onel$, and $\twol$ channels, respectively. The leptons considered in the analysis are electrons and muons.

The cross section measurement is performed in exclusive regions of phase space defined according to the type of vector boson, its transverse momentum $\ptV$, and the presence of additional jets. A standardized scheme for these measurement regions, called bins, is given by the definition of simplified template cross 
sections (STXS)~\cite{deFlorian:2016spz,Badger:2016bpw}, which were introduced to reduce the theoretical dependence in the measurements and to allow for a straightforward comparison of theoretical models with the measurements. 

While inclusive cross section measurements are suitable for establishing a given decay mode with limited sample size, using the STXS approach allows the kinematic properties of the Higgs boson production to be probed in a model-independent way. The sensitivity to the $\PV\PH$ production STXS bins is expected to be dominated by measurements of the $\PH\to\PQb\PAQb$ decay channel. An STXS measurement in this channel was performed by the ATLAS Collaboration with data collected between 2015 and 2018. The corresponding inclusive signal strength was measured to be $1.02^{+0.18}_{-0.17}$~\cite{ATLAS:2020fcp}.

This paper reports the study of $\PV\PH$ production with subsequent $\PH\to\PQb\PAQb$ decay, using the full CMS proton-proton ($\Pp\Pp$) collision data set collected in 2016--2018 at $\sqrt{s}=13\TeV$. The integrated luminosities are 36.3, 41.5, and 59.8 \fbinv for the 2016~\cite{CMS:2021xjt}, 2017~\cite{CMS:2018elu}, and 2018~\cite{CMS:2019jhq} data-taking periods, respectively, corresponding to a total of 138\fbinv.  Unlike in the previous publications~\cite{CMS:2013poe,CMS:2017odg,Sirunyan:2018kst}, dedicated categories are introduced for topologies that arise when the Higgs boson is highly Lorentz-boosted so that its decay products are reconstructed as a single merged jet. A similar large-radius-jet topology was explored by the ATLAS Collaboration~\cite{ATLAS:2020jwz}.

The paper is organized as follows. A brief description of the CMS detector and the simulated samples used in the analysis is given in Section~\ref{sec:sample}. The event reconstruction, with particular emphasis on the objects used in the measurement, is described in Section~\ref{sec:rec}. The trigger-level selection employed in the analysis is given in Section~\ref{sec:triggers}. The event selection and categorization, as well as the description of the STXS scheme, are documented in Section~\ref{sec:eventselection}. The sources of systematic uncertainties included in the measurement are listed in Section~\ref{sec:systematics}. The analysis strategy and results are discussed in Section~\ref{sec:results} with tabulated versions provided in HEPData~\cite{hepdata}. Section~\ref{sec:summa} provides the summary of the analysis results.

\section{The CMS detector and simulated samples}
\label{sec:sample}

The central feature of the CMS apparatus is a superconducting solenoid of 6\unit{m} internal diameter, providing a magnetic field of 3.8\unit{T}. Within the solenoid volume are a silicon pixel and strip tracker, a lead tungstate crystal electromagnetic calorimeter (ECAL), and a brass and scintillator hadron calorimeter (HCAL), each composed of a barrel and two endcap sections. Forward calorimeters extend the pseudorapidity ($\eta$) coverage provided by the barrel and endcap detectors. Muons are detected in gas-ionization chambers embedded in the steel flux-return yoke outside the solenoid. Events of interest are selected using a two-tiered trigger system~\cite{Khachatryan:2016bia}. The first level, composed of custom hardware processors, uses information from the calorimeters and muon detectors to select events at a rate of around 100\unit{kHz} within a fixed time interval of about 4\mus. The second level, known as the high-level trigger, consists of a farm of processors running a version of the full event reconstruction software optimized for fast processing, and reduces the event rate to around 1\unit{kHz} before data storage. A more detailed description of the CMS detector, together with a definition of the coordinate system used and the relevant kinematic variables, can be found in Ref.~\cite{Chatrchyan:2008zzk}.

Several Monte Carlo (MC) event generators are used to simulate signal and background processes. The signal processes contain Higgs bosons, with $m_\PH = 125 \GeV$, produced in association with \PW or \PZ bosons. In this analysis, only decays of the \PW and \PZ bosons involving muons, electrons, and/or neutrinos are considered. The quark-induced $\PZ\PH$ and $\PW\PH$ processes are generated at next-to-leading order (NLO) using the \POWHEG v2~\cite{Nason:2004rx,Frixione:2007vw,Alioli:2010xd} event generator extended with the MINLO procedure~\cite{Hamilton:2012np,Luisoni:2013cuh}, while the gluon-induced $\PZ\PH$ process (denoted $\Pg\Pg\PZ\PH$) is generated at leading-order (LO) accuracy with \POWHEG v2. The diboson processes $\PZ\PZ$, $\PW\PZ$, and $\PW\PW$ are simulated with \MGvATNLO~\cite{Alwall:2014hca} v2.2.2 (v2.4.2) in 2016 (2017 and 2018) simulations at NLO using the FxFx merging scheme~\cite{Frederix:2012ps} with up to two additional partons. For the analysis of 2017 and 2018 data, the same generator with NLO accuracy is used to simulate $\PW$+jets and $\PZ$+jets processes. For the analysis of 2016 data, LO accurate MadGraph simulations with the MLM matching scheme~\cite{Alwall:2007fs} are used to simulate $\PW$+jets and $\PZ$+jets events, in inclusive and \PQb-quark-enriched configurations. Corrections at NLO and next-to-NLO (NNLO) accuracy are applied to the 2016 LO $\PW$+jets and $\PZ$+jets event samples to achieve NNLO precision in shape and normalization. The NLO corrections are derived from simulation, while the NNLO corrections are obtained from theoretical calculations~\cite{nnlovjet}. While the NLO samples provide a more precise modeling of the kinematic properties of the $\PV$+jets processes, and are used for analyzing the 2017 and 2018 data sets, the corrections for the LO samples used in the analysis of the 2016 data are well validated. The production of simulated samples at NLO requires more computing resources than the production of samples at LO, which means the number of events in the NLO samples is limited. Thus, the LO samples are  used in the 2016 analysis to reduce the uncertainty in the measurement arising from the limited size of the simulated samples. NLO $\PV$+jets MC samples are produced in non-overlapping bins in jet multiplicity and $\ptV$, which are then merged to maximize the statistical power of the analysis. Two or more samples contributing to overlapping regions of the phase space are reweighted, such that the total cross section of a given process is conserved. The simulated $\PW$+jets and $\PZ$+jets samples at NLO are additionally corrected differentially in the angular momentum separation between jets ($\dRJJ$), in the region $\dRJJ < 1$. This correction is needed to account for a shape mismodeling observed in the region $\dRJJ < 1$. The correction improves the agreement between data and the NLO $\PV$+jets predictions. The reweighting is derived for $\PV$+jets events and is parametrized as flavor-agnostic, hence one single correction is derived for the $\PV$+jets processes, regardless of the jet flavor. This flavor-agnostic parametrization is justified because of the observation that the shape corrections are consistent across flavors of additional jets. The simulation-to-data reweighting is propagated to $\PV$+jets events in all analysis regions. Shape-altering uncertainties associated with this correction are uncorrelated across analysis channels, and will be discussed in Section~\ref{sec:systematics}.  

Samples of \ttbar events, as well as those of single top quark events produced in the $t$ channel, are generated with \POWHEG v2. Samples of single top quark events produced in the $\PQt\PW$ and $s$ channels are generated with \POWHEG v1.

The production cross sections used to normalize the simulated samples of signal and $\PV$+jets events are rescaled to NNLO quantum chromodynamic (QCD) predictions, inclusively in $\ptV$, and NLO EW accuracy, combining the results from Refs.~\cite{Ferrera:2013yga,Ferrera:2014lca,Brein:2012ne}, VH@NNLO~\cite{Brein:2012ne,Harlander:2013mla}, and HAWK v2.0~\cite{Denner:2014cla} generators, as described in Ref.~\cite{deFlorian:2016spz}. The NLO EW correction is applied as a function of $\ptV$. The production cross section for \ttbar events is calculated at NNLO with next-to-next-to-leading-logarithmic (NNLL) precision obtained using \TOPpp v2.0~\cite{Czakon:2011xx}. The parton distribution functions (PDFs) used to produce the NLO samples in the 2017--2018 analyses are from the NNLO NNPDF3.1 set~\cite{Ball:2014uwa}, while the LO NNPDF3.0 set is used for the LO samples. For parton showering and hadronization, all simulated samples are interfaced with \PYTHIA 8.2~\cite{Sjostrand:2007gs}. The \PYTHIA parameters for the underlying event description correspond to the CUETP8M1 tune for the samples compatible with the 2016 data set, and to the CP5 tune for the simulation corresponding to the 2017 and 2018 data sets. These tunes were derived in Ref.~\cite{Khachatryan:2015pea} based on the work described in Ref.~\cite{Skands:2014pea}. For all processes, the detector response is simulated with a detailed description of the CMS detector, based on the \GEANTfour~package~\cite{Agostinelli:2002hh}. The event reconstruction is performed with the same algorithms as for data. Additional interactions in the same or nearby bunch crossings, referred to as pileup (PU), are generated with \PYTHIA and added to the simulated samples. The simulated events are weighted such that the PU distribution in the simulation matches the one observed in data.

\section{Event reconstruction}
\label{sec:rec}
Events are reconstructed using a particle-flow (PF) algorithm~\cite{CMS:2017yfk}, which aims to reconstruct and identify each individual particle in an event (PF candidate) with an optimized combination of information from the various elements of the CMS detector. The energy of photons is obtained from the ECAL measurement. The energy of electrons is determined from a combination of the electron momentum at the primary interaction vertex as determined by the tracker, the energy of the corresponding ECAL cluster, and the energy sum of all bremsstrahlung photons spatially compatible with originating from the electron track. The energy of muons is obtained from the curvature of the corresponding track. The energy of charged hadrons is determined from a combination of their momentum measured in the tracker and the matching ECAL and HCAL energy deposits, corrected for the effect of hadronic showers on the calorimeter response. Finally, the energy of neutral hadrons is obtained from the corresponding corrected ECAL and HCAL energies. Events that are found to be affected by reconstruction failures and detector malfunctions are identified and rejected. Correction factors are applied to all reconstructed objects and used to equalize the reconstruction and the identification efficiencies in data and simulation.

The primary vertex (PV) is taken to be the vertex corresponding to the hardest scattering in the event, evaluated using tracking information alone, as described in Ref.~\cite{CMS-TDR-15-02}. The PV position is reconstructed using tracks clustered with the deterministic annealing algorithm~\cite{anneal}. The reconstructed PV is required to have a $z$ position within 24\unit{cm} of the nominal detector center, and a radial position within 2\unit{cm} of the beam axis. Displaced tracks originating from \PQb hadron decays are associated with secondary vertices. 

\PW and \PZ bosons are reconstructed using charged leptons and missing transverse momentum ($\ptvecmiss$). With two opposite-charge, same-flavor leptons, a full \PZ boson reconstruction is performed, which defines the $\twol$ channel. With one charged lepton and interpreting the missing transverse momentum in the event as the transverse momentum of a neutrino, the transverse momentum of the \PW boson candidate can be reconstructed, which defines the $\onel$ channel. In events without charged leptons, the large missing transverse momentum is used to estimate the $\Znn$ boson transverse momentum directly. This defines the $\zerol$ channel.

Electrons require the matching of a set of ECAL clusters, denoted as superclusters, to a track in the silicon tracker. The electron reconstruction is performed with the Gaussian sum filter algorithm~\cite{CMS:2020uim}. Electrons are preselected by requiring $\pt>7\GeV$, $\abs{\eta}<2.4$, $d_{xy}<0.05\unit{cm}$ and $d_z<0.2\unit{cm}$, where $d_{xy}$ and $d_z$ are the transverse and longitudinal impact parameters associated with the electron tracks, respectively.

A tighter identification is then performed using a multivariate approach (MVA ID). In addition, a set of offline requirements on ECAL-based electron quantities is applied. Two selections on the MVA ID discriminant are used, defining two different working points based on the expected electron identification efficiency of either $90\%$ (loose working point) or $80\%$ (tight working point). The loose working point is used when counting the number of additional leptons beyond the selected muons and electrons in each event, as well as for the event selection of the $\ZeeH$ channel. The tight working point is required to select events in the $\WenH$ channel. The electron \pt threshold in the $\WenH$ channel is $30\GeV$. For the $\ZeeH$ channel, the thresholds are 25 and $17\GeV$ for the two electrons. The working points and isolation requirements for the $\twol$ channel are generally looser than in the $\onel$ channel because requiring two leptons significantly reduces the background from QCD  multijet events. In the $\onel$ channel, tighter requirements are needed to reduce the multijet background. After applying the analysis selection, the QCD multijet background is found to be negligible.   

Muons are reconstructed from the combined fit of the tracker and muon detector signals~\cite{muCMS}. They are preselected by requiring the following conditions: $\pt>5\GeV$, ${\abs{\eta}<2.4}$, ${d_{xy}<0.5\unit{cm}}$, $d_z<1.0\unit{cm}$. Two working points corresponding to tight and loose muon identification requirements are utilized to reduce the fraction of other particles misidentified as muons. These working points depend on several of the following identification criteria: the number of hits in the tracker and muon system, the fit quality of the extrapolated muon track, and its consistency with the reconstructed PV. The muon \pt threshold in the $\WmnH$ channel is 25\GeV, and 25 and 15\GeV in the $\ZmmH$ channel.

The isolation of a lepton is defined relative to its momentum by summing the \pt of PF candidates, excluding the lepton itself, in geometrical cones around the lepton track direction at the event vertex. The cone size is expressed in terms of $\Delta R=\sqrt{\smash[b]{(\Delta\phi)^{2}+(\Delta\eta)^{2}}}$, where $\Delta\phi$ ($\Delta\eta)$ is the difference in the azimuthal angle (pseudorapidity) from the center of the cone to its edge. The lepton isolation criteria reject most of the major background consisting of nonprompt leptons produced in jets. The isolation cone for muons and electrons is $\Delta R=0.3$, and the ratio of the sum of each particle's \pt within the cone to the lepton \pt  must be smaller than 0.06.

Jets are reconstructed from PF candidates using the anti-\kt clustering algorithm with a distance parameter of 0.4 (AK4 jets)~\cite{CMS:2016lmd}. Jet momentum is determined as the vectorial sum of all particle momenta in the jet, and is found from simulation to be, on average, within 5--10\% of the true momentum over the entire \pt spectrum and detector acceptance. 

Pileup interactions can contribute additional tracks and calorimetric energy depositions to the event, increasing the apparent jet momentum. To mitigate this effect, tracks identified to be originating from PU vertices are discarded and an offset correction~\cite{CMS:2016lmd} is applied to correct for remaining contributions. 

Jet energy corrections are derived from simulation studies so that the average measured energy of jets is equal to that of particle-level jets. In situ measurements of the momentum balance in dijet, $\text{photon}$+$\text{jet}$, $\PZ$+$\text{jet}$, and multijet events are used to determine any residual differences between the jet energy scale (JES) in data and simulation, and appropriate corrections are made~\cite{CMS:2016lmd}. Additional selection criteria are applied to each jet to remove jets potentially dominated by instrumental effects or reconstruction failures. Jets that overlap geometrically ($\Delta R<0.4$) with preselected electrons or muons are discarded. Only jets with $\abs{\eta}<2.5$ are considered. In the $\PW\PH$ and $\ZnnH$ channels, a minimum threshold of jet $\pt > 25\GeV$ is used, while a looser selection $(\pt > 20\GeV)$ is applied in the $\ZllH$ channel. 

The jet energy resolution (JER) is about 15--20\% at 20\GeV, 10\% at 100\GeV, and 5\% at 1\TeV, and the jet energies in simulation are smeared to ensure their resolutions match those of jets in data~\cite{CMS:2016lmd}. 

Jets from final-state radiation (FSR), exceeding 20\GeV in momentum and fulfilling jet quality criteria are recovered by an FSR recovery algorithm, which adds the momenta of jets close to the Higgs boson candidate in the dijet mass calculation. The FSR recovery algorithm is applied to all analysis channels. 

A reweighting of the additional jet multiplicity spectrum in simulation is applied to the NLO $\PW$+jets and $\PZ$+jets samples used in the analysis of the 2017 and 2018 data sets. Here, additional jets are those jets that are retained in the analysis selection, but that do not stem from the Higgs boson candidate decay. This reweighting of the additional jet multiplicity is parametrized as a function of $\ptV$ and achieves improved modeling of the observed additional jet multiplicity distribution from simulation compared to data. Several shape-altering systematic uncertainties are associated with this reweighting, which will be discussed in Section~\ref{sec:systematics}. 

Jets that originate from the hadronization of \PQb quarks are identified by means of an algorithm based on a deep neural network (DNN), named \DeepCSV~\cite{btaggingDeepCSV}. This DNN has several probability outputs for jets resulting from quarks of different flavors. The algorithm provides a continuous discriminant output combining the information from track impact parameters and identified secondary vertices within jets, and from low-$\pt$ leptons produced by heavy-flavor quark decays present in the jet. A jet with a \DeepCSV discriminant value above a certain threshold is considered to be from the decay of a \PQb hadron, called a \PQb-tagged jet. The efficiency for tagging \PQb jets and the rate at which other jets are misidentified as \PQb jets depend on the chosen threshold of the \DeepCSV discriminant. The efficiency and the misidentification rate are both parametrized as functions of the jet \pt and $\eta$. The loose (tight) threshold has the highest (lowest) efficiency for tagging \PQb jets, while allowing the most (least) contamination from light, i.e., \PQu, \PQd, \PQs, and gluon (\Pg) jets, as well as \PQc jets. The working points are defined such that a specific target for the misidentification (mistag) rate is achieved: 10, 1.0, and 0.1\% for the loose, medium, and tight working points, respectively. Each channel is optimized separately, and results in the same selection: the leading \PQb jet candidate in \PQb tagging discriminant score must pass the medium identification working point of the \DeepCSV algorithm. For the next-to-highest \PQb tagging discriminant score (subleading) \PQb jet candidate, the optimal selection requires the loose working point of the \DeepCSV discriminant. Along with these two \PQb-tagged jets associated with the Higgs boson candidate reconstruction, additional \PQb-tagged jets can be present in the event. These jets are included in the analysis as discussed in Section~\ref{sec:eventselection}. The \PQb tagging selections for all three channels will be discussed in more detail in Section~\ref{sec:eventselection}. 

Some \PQb tagging inputs are used in the training of the multivariate discriminants employed in the analysis signal region (SR) to separate signal from background events. The \PQb tagging discriminant is corrected to equalize the efficiency in data and simulation as a function of the \PQb tagging score and the kinematic properties of each jet. An index with an integer value between 0 and 3 referring to the \PQb tagging working point requirement that the jet fulfills is used as one of the inputs for deriving the signal-versus-background discriminants.

The dijet invariant mass resolution is computed by applying a multivariate regression analysis using a DNN trained on simulated \PQb jets stemming from \ttbar events~\cite{breg}. The training includes input features that describe the jet energy and direction, as well as properties of the secondary vertices of the jets. Information about tracks associated with jets, jet constituents, low-$\pt$ electrons and muons in the jet associated with semileptonic \PB hadron decays is also used. The \PQb jet energy regression improves the precision of the jet four-vectors, which leads to a 10--15\% improvement in the dijet invariant mass resolution, depending on the \pt of the reconstructed Higgs boson candidate. The momenta of the Higgs boson candidate jets are corrected by the application of the \PQb jet energy regression described above, unless otherwise specified. After the application of the regression, a dedicated smearing is applied to the \PQb jet energy so that the dijet invariant mass resolution in simulation will match the performance in data. The smearing parameters are extracted in events where a jet recoils against a \PZ boson that decays into leptons. Because the \PZ boson \pt is balanced with the jet $\pt$, and given that the lepton momentum measurement is precise, the ratio of the reconstructed jet momentum ($\pt^{j}$) to the \PZ boson momentum ($\pt^{\Pell\Pell}$) enables a precise measurement of the jet momentum and energy. The distribution of the energy difference between the \pt balance procedure and the \PQb jet energy regression is used to estimate the \PQb jet energy regression scale and resolution corrections and uncertainties. The selected events are divided into four regions of $\alpha=\pt^{j}/\pt^{\Pell\Pell}$, and this procedure is applied in each $\alpha$ bin.

In signal events where the Higgs boson has a \pt exceeding 250\GeV, two jets reconstructed with the AK4 clustering~\cite{Cacciari:2008gp} algorithm will begin to overlap as the opening angle between the jets shrinks. Therefore, these events are reconstructed using a single large radius of 0.8, producing what are termed AK8 jets. The modified mass drop tagger algorithm (soft-drop algorithm)~\cite{softd} is applied to remove soft and wide-angle radiation. This algorithm identifies two hard AK4 subjets within the AK8 jet. The mass of the AK8 jet, upon application of the soft-drop algorithm ($m_{\mathrm{SD}}$), is used as discriminating variable in the analysis. The four-momenta of the two subjets are used to calculate the kinematic properties of their corresponding AK8 jet.

The \DeepAKviii algorithm~\cite{deepak8algo} is used to tag boosted $\PH\to\PQb\PAQb$ topologies, exploiting AK8 jet (${\Delta R=0.8}$) reconstruction. The tagger architecture is based on a set of convolutional kernels spanning multiple candidates. It makes use of a multioutput feed-forward neural network with low-level input features (PF candidates), in addition to the traditional observables of the boosted-jet environment (tracks, jets, and secondary vertices). 

Ten features for each charged and neutral PF candidate are passed to one of these convolutional kernels, ordered in candidate momentum, to learn the jet substructure. The flavor content of the jet is learned by two other kernels. One of these uses only charged constituents of the jet, sorted by the displacement with respect to the PV, while the other one uses secondary vertices. The \DeepAKviii algorithm is decorrelated from the mass of the AK8 jet, which is included in the training of the network. The level of decorrelation is such that no accumulation of events in mass is observed even for very small values of the background mistag rate. This means that the AK8 jet mass shape associated with the background does not become similar to that of the signal after selection with the \DeepAKviii tagger. The \DeepAKviii algorithm aims to classify a variety of resonances in multiple decay modes. It provides separation for the Lorentz-boosted $\PH\to\PQb\PAQb$ signal against several background outputs: merged light jets, top quark jets, and QCD multijets. Scale factors are used to correct the simulation to account for differences in efficiency with respect to the data. These scale factor are measured in highly-boosted gluon splitting into $\PQb\PAQb$ events. They are applied to the $\PH\to\PQb\PAQb$ boosted signal output node. The scale factors are parametrized as a function of the AK8 jet \pt and $\eta$, and are available for two working points, as discussed in Section~\ref{sec:eventselection}.

The value of $\ptvecmiss$ is crucial in the reconstruction of the $\WtoLN$ and $\ZtoNN$ decays. The $\ptvecmiss$ is computed as the negative vector $\ptvec$ sum of all the PF candidates in an event, and its magnitude is denoted as $\ptM$~\cite{CMS:2019ctu}. The calculation of $\ptvecmiss$ is modified to account for corrections to the energy scale of the reconstructed jets in the event. Track-based missing transverse momentum, denoted as $\ptMtrk$, is also used in the analysis. Only tracks that have a \pt above a minimum momentum threshold (2 \GeV) and an impact parameter consistent with the PV are considered in the vectorial sum. The tracks are also required to pass quality requirements, which were designed to limit the contribution from misreconstructed tracks. The estimation of $\ptvecmiss$ in simulated events is improved by correcting it for the difference between raw and calibrated jets, including scale and resolution corrections. In addition, a set of recommended filters to remove known issues related to instrumental noise and problematic events is applied~\cite{CMS:2019ctu}.

In background event topologies, such as $\ttbar$ events, there is additional low-energy jet activity in the event (soft jets). A collection of additional tracks (or PF candidates) is built by requiring $\pt>300\MeV$, $d_z<2\unit{mm}$, a ``high-purity'' quality identification, and not be associated with either the leptons from the vector boson decay nor with the two selected \PQb-tagged jets in the event. A collection of soft jets is clustered from these tracks using the anti-\kt clustering algorithm with a distance parameter of $\Delta R<0.4$. The soft activity with a $\pt$ threshold at 5 \GeV is used as a discriminant variable for analysis channels where its modeling in simulation is satisfactory, both in the resolved and boosted topologies, as discussed in Section~\ref{sec:eventselection}.

\section{Trigger-level selection}
\label{sec:triggers}
Several triggers are used to collect events containing final-state particles consistent with the signal processes considered. The trigger selection
focuses on the final state of the \PW or \PZ boson produced in association with the Higgs boson. The triggers used to select events in the 0-lepton channel make use of \ptmiss and missing hadronic transverse momentum, \mht. These quantities are derived from the reconstructed objects as identified by the PF algorithm. Online, \mht is defined as the magnitude of the negative vector $\ptvec$ sum of all reconstructed jets with $\pt>20\GeV$ and $\abs{\eta}<2.5$. The main triggers used in each of the data-taking periods require the same threshold on $\ptmiss$ and $\mht$. This threshold is 110\GeV in 2016, and 120\GeV in the 2017 and 2018 data-taking periods. In the 1-lepton channel, single-lepton triggers are used. The \pt threshold for electrons is 27\GeV in the 2016 data-taking period, rising to 32\GeV in 2017--2018. For muons, the \pt threshold is 24\GeV in the 2016 and 2018 data-taking periods, and is increased to 27\GeV in 2017. Dilepton triggers are used to select events in the 2-lepton channel. The \pt thresholds for electrons are 23 and 12\GeV in all data-taking periods. For muons, the \pt thresholds are 17 and 8\GeV in all data-taking periods; the triggers used in 2017 and 2018 differ from those used in 2016 by the additional requirement that the dimuon invariant mass must be greater than 3.8\GeV. In addition to the \pt thresholds, the triggers require the leptons to pass stringent identification criteria. The trigger-level leptons are also required to be isolated from other tracks and energy deposits in the calorimeters.

\section{Event selection and multivariate discriminants for signal extraction}
\label{sec:eventselection}

Though there are numerous features that differ between the three lepton channels and the $\PQb\PAQb$ reconstruction topologies, a unified analysis strategy is used.

Three control regions (CRs), each enriched in one of the three primary backgrounds: $\ttbar$, $\PV$+light-quark jets, and $\PV+\PQb$ jets, are defined for every channel and \PQb quark reconstruction topology. The jet flavor in the $\PV$+jets processes ($\PV+\PQb\PQb$ jets, $\PV+\PQb$ jets, $\PV+\PQc$ jets and$\PV$+light-quark jets) is defined based on the presence of the corresponding hadrons at generator-level with transverse momentum exceeding 25 \GeV and $\abs{\eta}<2.5$. Templates derived from simulation are fitted to the data with the normalizations of the three primary backgrounds left unconstrained and incorporating several systematic uncertainties that are allowed to modify the shapes. The SRs are defined by requiring the dijet invariant mass to be in the range 90--150\GeV and making a relatively tight selection on the multivariate quark flavor discriminant that target two \PQb quarks. The selection values on the multivariate quark flavor discriminants for the SR definitions are reported in Tables~\ref{tab:0LepSelection}, \ref{tab:1LepSelection} and \ref{tab:2LepSelection} for the $\zerol$, $\onel$ and $\twol$ channels, respectively. The fitted observables in the SRs are all DNN outputs, binned such that all bins contain roughly equal expected signal yields. The observables used in the CRs vary between the channels, and are selected to constrain particular features of the background model. All SRs and CRs are fitted simultaneously to extract the background shapes, background normalizations, and signal strengths.

The requirement of an identified boosted \PW or \PZ bosons in the signal events suppresses backgrounds from QCD multijet events, while also providing an efficient trigger path when the \PW or \PZ boson decays to charged leptons. Requiring a large boost provides additional advantages. It further reduces the large backgrounds from \PW and \PZ production in association with jets, helps to suppress the large background from top quark production in the signal channels including neutrinos, and generally improves the invariant mass resolution of the reconstructed Higgs boson candidates. Therefore, in addition to the analysis targeting \textit{resolved} events where the \PQb jets from the decay of the Higgs boson candidate are reconstructed as separate AK4 jets, we also include \textit{boosted} events where the \PQb jets from the decay of the Higgs boson candidates are both contained in a large-radius jet reconstructed with the AK8 algorithm. 

The boosted analysis only considers events with $\ptV>250\GeV$. The SRs and CRs for these analyses are optimized separately for the three analysis channels (0-, 1-, and 2-lepton). Sections~\ref{sec:resolved} and~\ref{sec:boosted} describe the main features of the resolved and boosted analyses. 

Some of the events with $\ptV>250\GeV$ can be reconstructed in both the resolved and boosted topologies. We refer to these as overlap events in what follows. There are four categories that overlap events can enter, depending on whether they are resolved or boosted, and whether they belong to the SRs or the CRs. By studying all possible permutations in simulation, the following priority ranking was selected: resolved SRs, boosted SRs, resolved CRs, and boosted CRs. The reason for this choice is that the resolved SRs provide more sensitivity, gauged as the uncertainty in the signal strength extracted using an Asimov dataset, to the signal than the boosted SRs. Overlap events that would be placed either in a resolved CR or a boosted SR are assigned to the boosted SR to improve the sensitivity of the analysis. This ranking minimizes the expected uncertainties in the STXS measurements and ensures that events satisfying both resolved and boosted selection criteria are used only once.

\subsection{Subcategorization in STXS bins}
\label{sec:resolved_stxscats}

\begin{figure}[!htb]
    \centering
    \includegraphics[width=0.5\textwidth]{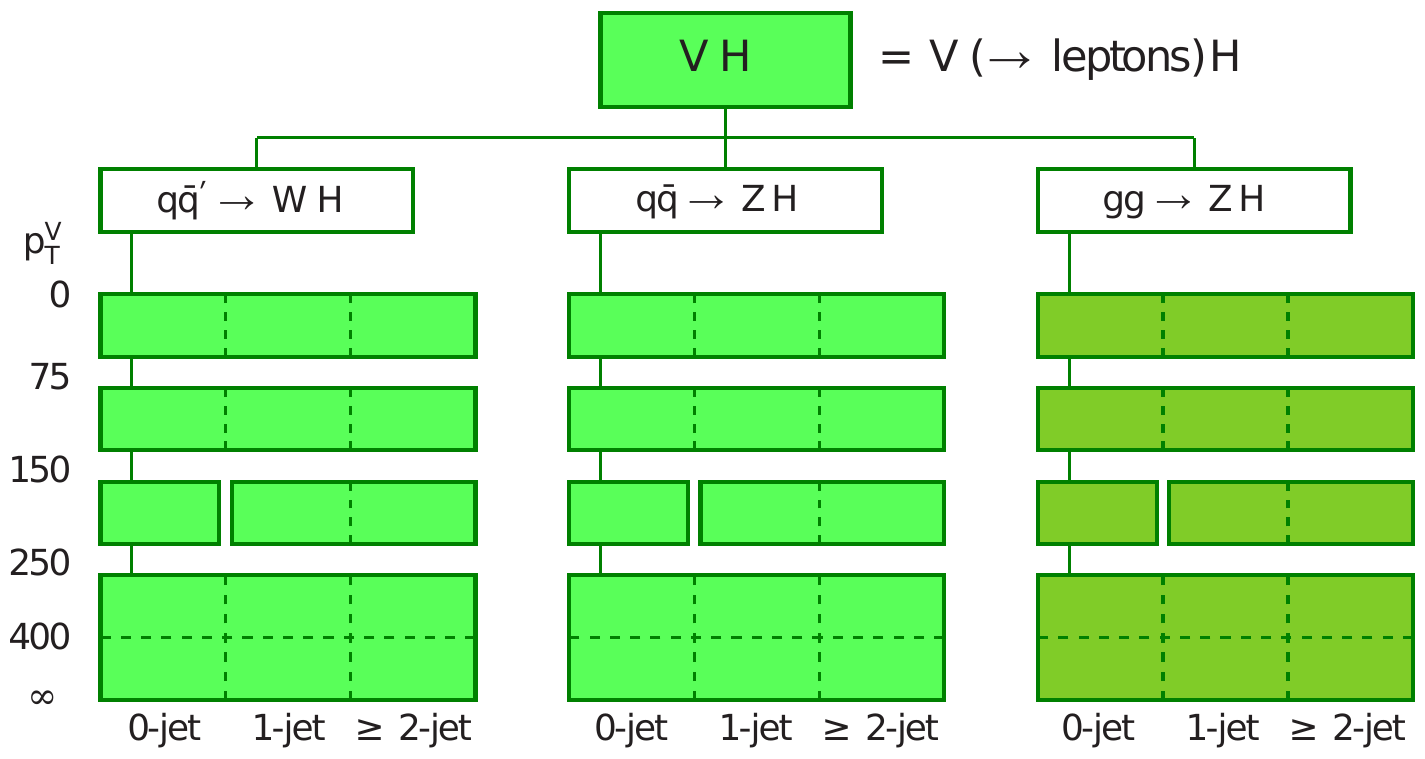}
    \caption{Overview of the STXS bins for the three $\PV\PH$ production modes~\cite{Badger:2016bpw}. The vertical axis reflects the $\ptV$ bin ranges and the horizontal axis the number of additional jets. The general bin definitions are indicated by the green boxes. No distinction is made between gluon- and quark-induced production modes in the analysis. As mentioned in Section~\ref{sec:resolved_stxscats}, some STXS bins are not explicitly targeted by the analysis: contributions from these bins are fixed to their SM expectations.}
    \label{fig:stxs}
\end{figure}

In all channels, additional subcategorizations are employed to maximize the signal sensitivity to the different STXS bins. The STXS categorization employed in the analysis is shown in Fig.~\ref{fig:stxs}. The STXS binning for the V(leptonic)H process~\cite{Badger:2016bpw} uses a division into three production modes: $\PW\PH$, $\PQq\PQq\to \PZ\PH$, and $\Pg\Pg\to \PZ\PH$. For each of these production modes, the STXS classification defines bins in $\ptV$: ${[0,75)}$, ${[75,150)}$, ${[150,250)}$,  ${[250,400)}$,  ${>400\GeV}$. 
 
The $[150,250)\GeV$ $\ptV$ STXS bin is split into two; one bin without any additional jets and another bin with additional jets. This categorization constitutes the STXS classification, that is, the generator-level division of events into STXS bins. Not all of these bins are accessible in the analysis. The gluon- and quark-induced $\PZ\PH$ production modes are merged because the sensitivity to the separate processes is small with the currently available data set. In addition, the two exclusive jet bins for the $\PW\PH$ process are merged because of the low sensitivity to the bin with additional jets. Apart from bins that are merged, there are also STXS bins that are not within the analysis selection acceptance. This concerns the lowest $\ptV$ bin, $[0,75)\GeV$, for all production processes, and the $[75,150)\GeV$ bin for the $\PW\PH$ process. Negligible contributions from these bins can appear in the analysis because generator-level quantities used to define the STXS bins do not exactly match their reconstruction-level equivalents. Where these bins do contribute to the analysis, their rates are fixed to the SM expectations. To target the STXS bins, we define corresponding reconstruction-level categories, constituting the subcategorization of the five channels considered in the measurement. 
\begin{itemize}
\item For the $\zerol$ channel, we define three categories in reconstructed $\ptV$: ${[75,250]\GeV}$, ${[250,400)\GeV}$, and ${>400\GeV}$.
\item For the $\twol$ channel, we define four categories in reconstructed $\ptV$: ${[75,150)\GeV}$, ${[150,250)\GeV}$, ${[250,400)\GeV}$, and ${>400\GeV}$. 
\item In the 0- and 2-lepton channels, the ${150 < \ptV < 250\GeV}$ category is further subdivided into a subcategory without any additional jets and another with at least one additional jet. 
\item For the $\onel$ channel, we define three categories in reconstructed $\ptV$: ${[150,250)\GeV}$, ${[250,400)\GeV}$, and ${>400\GeV}$. There is no further subdivision into additional jet categories.
\end{itemize}

\subsection{Analysis of the resolved jet topology}
\label{sec:resolved}

\subsubsection{Analysis of the 0-lepton channel}
The event topology of the $\zerol$ channel is characterized by the presence of large $\ptM$ due to the $\ZtoNN$ decay and a pair of \PQb jets from the Higgs boson decay recoiling against the \PZ. Additional jet activity ($\Naj$) is expected to be low, and no high-$\pt$ leptons should be present. Events in the $\zerol$ channel are selected using the high-level trigger paths described in Section~\ref{sec:triggers}. The $\ptM$ is required to be larger than 170\GeV. As a result of the trigger acceptance, we require $\minMETMHT$ to exceed 100\GeV. The Higgs boson candidate is reconstructed using the two jets with $\abs{\eta}<2.5$ that have the highest \PQb tagging score. The leading and subleading \PQb jet associated with the Higgs boson candidate, denoted with the suffixes $j_{\mathrm{1}}$ and $j_{\mathrm{2}}$, must have $\pt>60$ and 35\GeV, respectively. Since the Higgs boson recoils against the \PZ boson, we require the difference in azimuthal angle between the $\jj$ system and the \PZ boson to satisfy $\dphiZjj$ $>2.0$. We also demand the invariant mass of the dijet system to be in the range 50--500\GeV and have $\pt>120\GeV$. Events containing at least one isolated lepton with ${\pt>15\GeV}$ in the central region ($\etaTF$) are rejected. To reduce the QCD multijet background, a requirement of $\dphiMJ>0.5$ is applied for all jets with ${\pt>30\GeV}$ for both the SR and CRs. This means that events with any energetic jets close to $\ptvecmiss$ in azimuth are not considered. Events are split into four orthogonal categories, one SR and three CRs. The selections used for the SR, as well as for the CRs enriched in the main background processes ($\PZ$+$\PQb$ jets, $\PZ$+light-quark jets, and $\ttbar$), are summarized in Table~\ref{tab:0LepSelection}. The notation $\text{btag}(\mathrm{j_{1}})$ ($\text{btag}(\mathrm{j_{2}})$) $>$ ($<$) medium (loose) indicates that the \PQb tagging score associated with the leading (subleading) jet of the Higgs boson candidate is required to be larger (smaller) than the medium (loose) \PQb tagging working point. 

\begin{table*}[!htb]
\centering
\topcaption{Definition of the SR and CRs for the resolved selection in the $\zerol$ channel. If the same selection is applied in all SRs and CRs, this is indicated by the $\div$ symbol in the latter. If no selection is applied, this is indicated by the $\NA$ symbol. The $\Mjj$ and momenta variables have units of \GeV.}
\begin{scotch}{lcccc}
Variable & SR & $\PZ$+$\PQb$ jets & $\PZ$+light-quark jets & $\ttbar$ \\
\hline
Common selection: & & & &\\ [\cmsTabSkip]
$\minMETMHT$ & $>$100 & $\div$ & $\div$ & $\div$\\
$\ptM$ & $>$170 & $\div$ & $\div$ & $\div$\\
$\pt(\mathrm{j_1})$ & $>$60 & $\div$ & $\div$ & $\div$\\
$\pt(\mathrm{j_2})$ & $>$35 & $\div$ & $\div$ & $\div$\\
$\ptjj$& $>$120 & $\div$ & $\div$ & $\div$\\
$\dphiZjj$ & $>$2.0 & $\div$ & $\div$ & $\div$ \\
$\dphiMJ$ & $>$0.5 & $\div$ & $\div$ & $\div$ \\ [\cmsTabSkip] 
SR/CR difference: & & & &\\  [\cmsTabSkip] 
$\Naj$& $<$2 & $<$2 & $<$2 & $\geq$2 \\
$\Mjj$& $\in$[90--150]& $\notin$[90--150] and $<$250 & $<$250  & $<$250 \\
$\text{btag}(\mathrm{j_1})$& $>$medium & $>$medium & $<$medium & $>$medium \\ 
$\text{btag}(\mathrm{j_2})$ & $>$loose  & $>$loose & $<$loose & $>$loose\\  
$\dphiMtkM$ & $<$0.5 & $<$0.5 &  $<$0.5 & \NA \\
$\min{\dphiMJ}$ & \NA & \NA & \NA & $<\pi/2$\\
\end{scotch}
\label{tab:0LepSelection}
\end{table*}

\subsubsection{Analysis of the 1-lepton channel}
The topology of the $\onel$ events is characterized by the presence of a single isolated lepton from the decay of the \PW boson recoiling against two \PQb jets from the decay of the Higgs boson. The presence of a single isolated lepton provides a trigger path for this channel. The Higgs boson is reconstructed using the two jets, with $\pt>25\GeV$, that have the highest \PQb tag scores, denoted with the suffixes $j_{\mathrm{1}}$ and $j_{\mathrm{2}}$. Only jets with $\abs{\eta}<2.5$ are considered. We require the Higgs and \PW boson candidates to have $\pt > 100$ and 150 \GeV, respectively. An additional requirement on the ratio of \ptmiss to its uncertainty, $\ptM/\sigma(\ptM)$, is applied to the CRs enriched in $\PW$+$\PQb$ and $\PW$+light-quark jets as well as to the SR. Events are not considered if they contain additional leptons ($N_{\text{al}}$) with $\abs{\eta}<2.5$ and $\pt>25\GeV$. The selected events are split into four orthogonal categories, one SR and three CRs. The selections used for the SR, as well as for the CRs enriched in the main background processes ($\PW$+$\PQb$ jets, $\PW$+light-quark jets, and $\ttbar$), are summarized in Table~\ref{tab:1LepSelection}. The notation $\text{btag}(\mathrm{j_{1}})$ ($\text{btag}(\mathrm{j_{2}})$) $>$ ($<$) loose, medium, or tight indicates that the \PQb tagging score associated with the leading (subleading) jet of the Higgs boson candidate is required to be larger (smaller) than the corresponding loose, medium, or tight \PQb tagging working point. The notation [loose-medium] indicates that the leading jet associated with the Higgs boson candidate is required to have a \PQb tagging score in the range loose--medium \PQb tagging working point. 
 
\begin{table*}[!htb]
\centering
\topcaption{Definition of the SR and CRs for the resolved selection of the $\onel$ channel. If the same selection is applied in all SRs and CRs, this is indicated by the $\div$ symbol in the latter. If no selection is applied, this is indicated by the $\NA$ symbol. The $\Mjj$ and momenta variables have units of \GeV.}
\begin{scotch}{lcccc}
Variable & SR & $\PW$+$\PQb$ jets & $\PW$+light-quark jets & $\ttbar$ \\
\hline
Common selection: & & & &\\  [\cmsTabSkip] 
$\ptjj$ & $>$100 & $\div$ & $\div$ & $\div$\\ 
$\pt(\PV)$ & $>$150 & $\div$ & $\div$ & $\div$\\ 
$N_{\text{al}}$ & $<$1 & $\div$ & $\div$ & $\div$\\ 
$\pt(\mathrm{j_1})$ & $>$25 & $\div$ & $\div$ & $\div$\\ 
$\pt(\mathrm{j_2})$ & $>$25 & $\div$ & $\div$ & $\div$\\ 
$\Delta\phi(\text{lep}, \ptvecmiss)$ & $<$2 & $\div$ & $\div$ & $\div$\\  [\cmsTabSkip] 
SR/CR difference: & & & &\\  [\cmsTabSkip] 
$\text{btag}(\mathrm{j_1})$& $>$medium & $>$medium & [loose-medium] &$>$tight\\
$\text{btag}(\mathrm{j_2})$& $>$loose & $>$loose & \NA & \NA \\
$\Mjj$ & $\in$[90--150] & $\in$[150--250] and $<$90 & $<$250 & $<$250\\
$\Naj$ & $<$2 & $<$2 & \NA & $\geq$2 \\
$\frac{\ptM}{\sigma(\ptM)}$ & $>$2 & $>$2 & $>$2 & \NA \\
$\dphiWjj $ & $>$2.5 & $>$2.5 & \NA & \NA\\
\end{scotch}
\label{tab:1LepSelection}
\end{table*}

\subsubsection{Analysis of the 2-lepton channel}

\begin{figure}[!htb]  
\centering 
\includegraphics[width=0.48\textwidth]{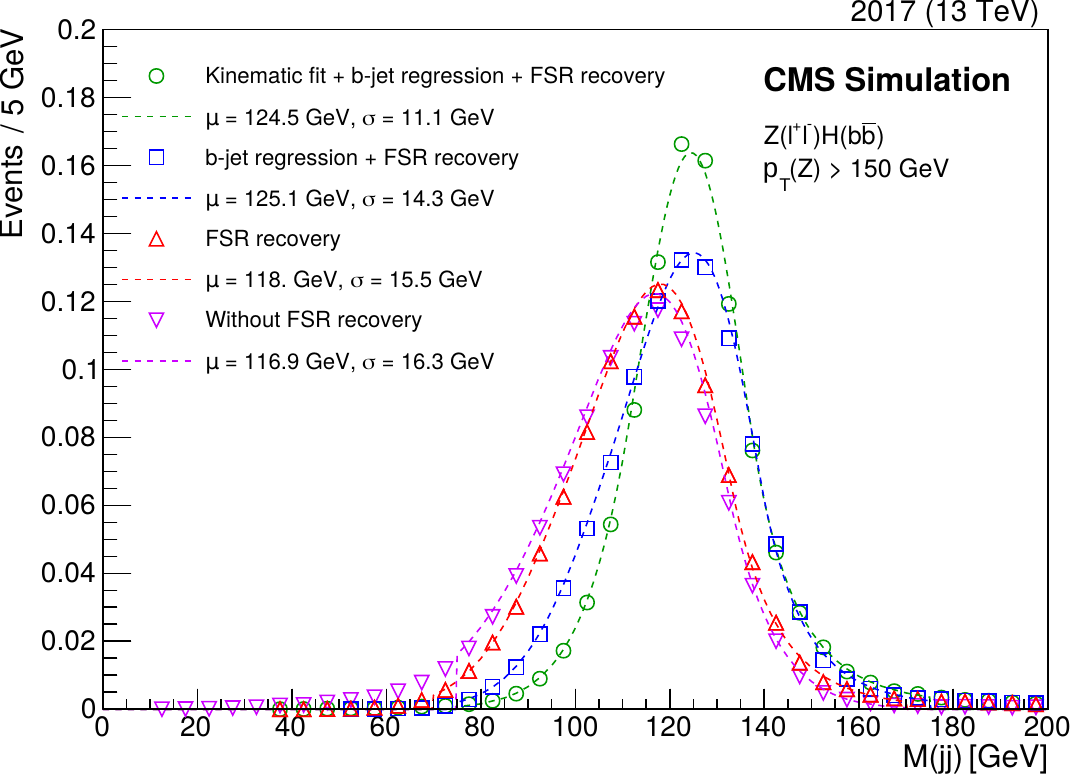} 
\caption{Dijet invariant mass distributions in samples of simulated (2017 simulation) signal events passing the $\twol$ channel requirements without any additional recoiling jet. Distributions are shown without the usage of the FSR recovery algorithm (purple triangles), before (red triangles) and after (blue squares) the energy corrections from the \PQb jet regression are applied, and when a kinematic fit procedure (green circles) is used in addition to them. The fitted mean and width of the core of the distribution, obtained by fitting a Bukin function~\cite{bukin}, are displayed in the figure. The statistical uncertainties are smaller than the marker height.}
\label{fig:mjjplot_res}
\end{figure}

The topology of the signal events in the $\twol$ channel is characterized by the presence of two isolated leptons from the decay of the \PZ boson, recoiling against two \PQb jets from the decay of the Higgs boson. Figure~\ref{fig:mjjplot_res} shows the simulated invariant mass for signal events (without any additional recoiling jet) generated with $m_{\PH}=125\GeV$ in the $\twol$ channel using the FSR recovery algorithm mentioned in Section~\ref{sec:rec}, along with the addition of the \PQb jet energy regression (discussed in Section~\ref{sec:rec}), and a kinematic fit. Since there is no genuine $\ptvecmiss$ in the hard-scattering process for this event topology, a kinematic fit is performed to improve the resolution of the dijet invariant mass. In this fit, the mass of the dilepton system is constrained to the \PZ boson mass (PDG), while the total \pt of all the particles must sum to zero. This kinematic fit imposes a balance between the \PZ boson $\pt$, which is expected to be well measured because of the good momentum resolution for high-$\pt$ leptons, and the vectorial sum of the jet momenta. The objects used in the fit are the two \PQb-tagged jets that form the Higgs boson candidate after application of the \PQb jet energy regression, and the FSR recovery algorithm, two lepton candidates, and up to one jet produced by initial-state radiation (ISR). Only jets with $\abs{\eta}<2.5$ are included in the analysis. As shown in Fig.~\ref{fig:mjjplot_res}, a large improvement in resolution in the region of the mass peak is achieved when the kinematic fit, the \PQb jet energy regression, and the FSR recovery algorithms are employed together.

The Higgs boson is reconstructed using the two highest \PQb tagging score jets, denoted with the suffixes $j_{\mathrm{1}}$ and $j_{\mathrm{2}}$, that have $\pt>20\GeV$, and $\ptV$ is required to exceed 75\GeV. The selections used to define the SR and CRs, which are enriched in the main background processes ($\PZ$+$\PQb$ jets, $\PZ$+light-quark jets, $\ttbar$) are summarized in Table~\ref{tab:2LepSelection}. Similarly to the 0- and 1-lepton channels, the $\ttbar$ CR is very pure in $\ttbar$ events. The additional requirement that the dilepton invariant mass lie in the \PZ boson mass window ensures the $\PZ$+jets purity in the corresponding regions is high. The notation $\text{btag}(\mathrm{j_{1}})$ ($\text{btag}(\mathrm{j_{2}})$) $>$ ($<$) loose, medium, or tight indicates that the \PQb tagging score associated with the leading (subleading) jet of the Higgs boson candidate is required to be larger (smaller) than the corresponding loose, medium, or tight \PQb tagging working point. 

\begin{table*}[!htb]
\centering
\topcaption{Definition of the SR and CRs for the resolved selection in the $\twol$ channel. If the same selection is applied in all SRs and CRs, this is indicated by the $\div$ symbol in the latter. If no selection is applied, this is indicated by the $\NA$ symbol. The $\Mjj$, $M(\PV)$, and momenta variables have units of \GeV.} 
\begin{scotch}{lcccc}
Variable & SR & $\PZ$+$\PQb$ jets & $\PZ$+light-quark jets & $\ttbar$ \\
\hline
$\pt(\PV)$ & $>$75 & $\div$ & $\div$ & $\div$\\ 
$\text{btag}(\mathrm{j_1})$& $>$medium & $>$medium & $<$loose & $>$tight\\ 
$\text{btag}(\mathrm{j_2})$& $>$loose  & $>$loose & $<$loose & $>$loose\\  
$M(\PV)$ & $\in$[75--105] & $\in$[85--97] & $\in$[75--105] & $\in$[10--75] and $>$120\\ 
$\Mjj$ & $\in$[90--150] & $\notin$[90--150] and $<$250 & $\in$[90--150] & $<$250\\ 
$\dphiZjj $ & $>$2.5 & $>$2.5 & $>$2.5 & \NA\\
\end{scotch}
\label{tab:2LepSelection}
\end{table*}

\subsubsection{Multivariate discriminants in the SR and the heavy-flavor CRs}
\label{sec:resolvednn}
In order to improve the separation between signal and background, a DNN classifier trained to discriminate the $\PV\PH$, $\PH\to\PQb\PAQb$ signal against all background processes is used for all the channels. The same network architecture is used for both the DNN signal/background classifier and the DNN multiclass background classifier discussed later in this section. A DNN classifier with six hidden fully connected layers is trained, with each layer containing 512, 256, 128, 64, 64, and 64 nodes~\cite{deepl}. The final layer is a softmax layer~\cite{deepl}, giving the probability for an event to be in a particular class. The classifier is trained to minimize the cross-entropy loss function~\cite{deepl} using the Adam optimizer algorithm~\cite{adamdnn} trained on a minibatch of size 1024.

The input features used in the DNN training encompass the kinematic properties of the final state: the masses, momenta, and angles of the jets, dijet, vector boson candidate, and leptons. The additional reconstructed jet multiplicity is also used. While many potentially discriminating variables were considered, variables that did not contribute to the analysis sensitivity were dropped. The modeling of these variables in the simulation is also inspected. When the modeling of such variables is observed to be unsatisfactory, the variables are removed from the input feature list of the DNN training. 

The full list of input variables is shown in Table~\ref{tab:DNNinputs_list} for the 0-, 1-, and 2-lepton channels. These variables are used as inputs to the DNN trainings in all the STXS categories in which the DNN trainings are performed separately. To evaluate the recoil jet multiplicity, used as an input variable in the DNN training for the 2-lepton channel, recoil jets are identified as ISR jets. They must have ${\pt>20\GeV}$, pass reconstruction quality requirements, and cannot be part of the Higgs boson candidate or one of the FSR jets. 

\begin{figure}[!htb]
\centering 
\includegraphics[width=0.4\textwidth]{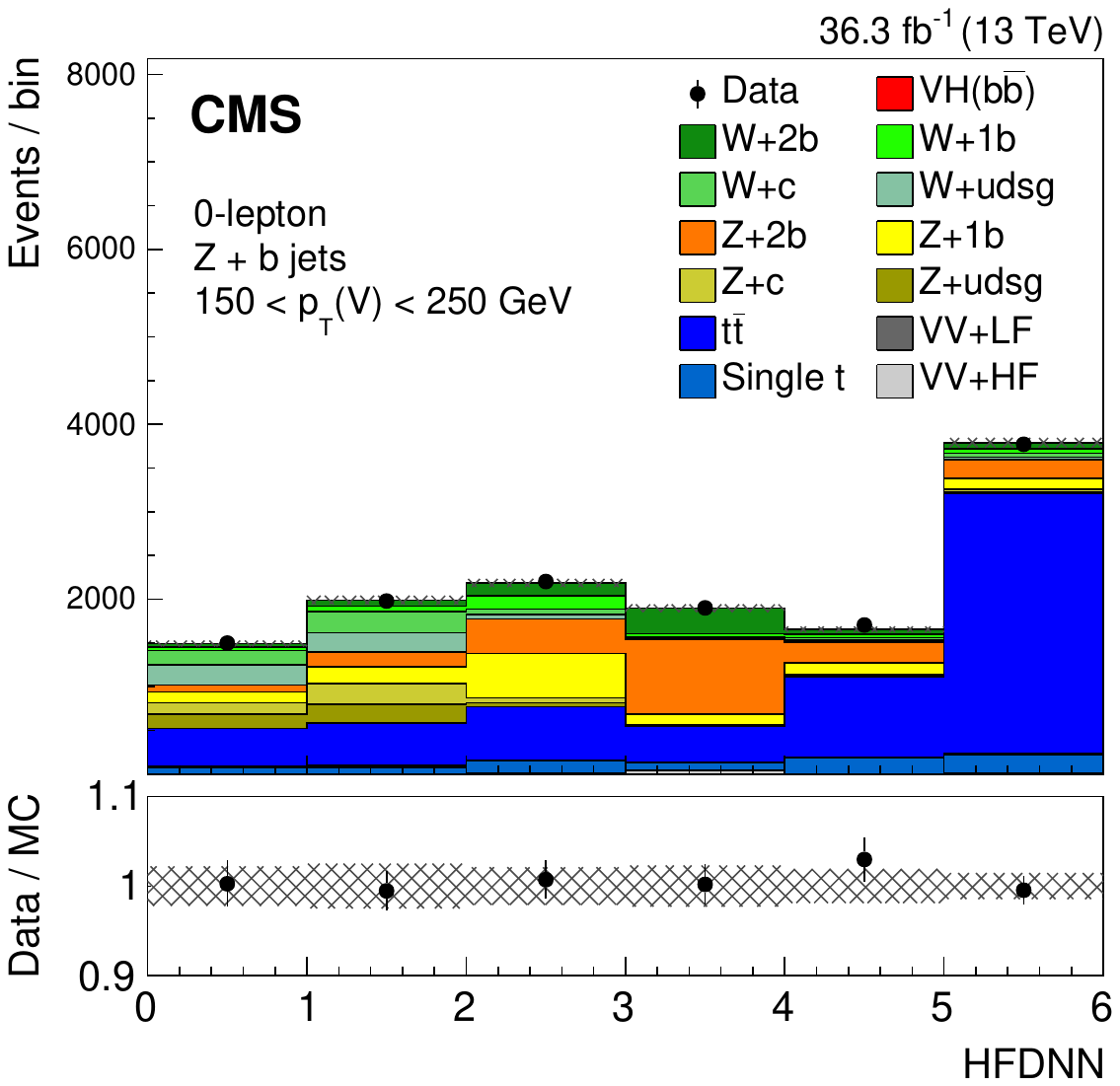} 
\includegraphics[width=0.4\textwidth]{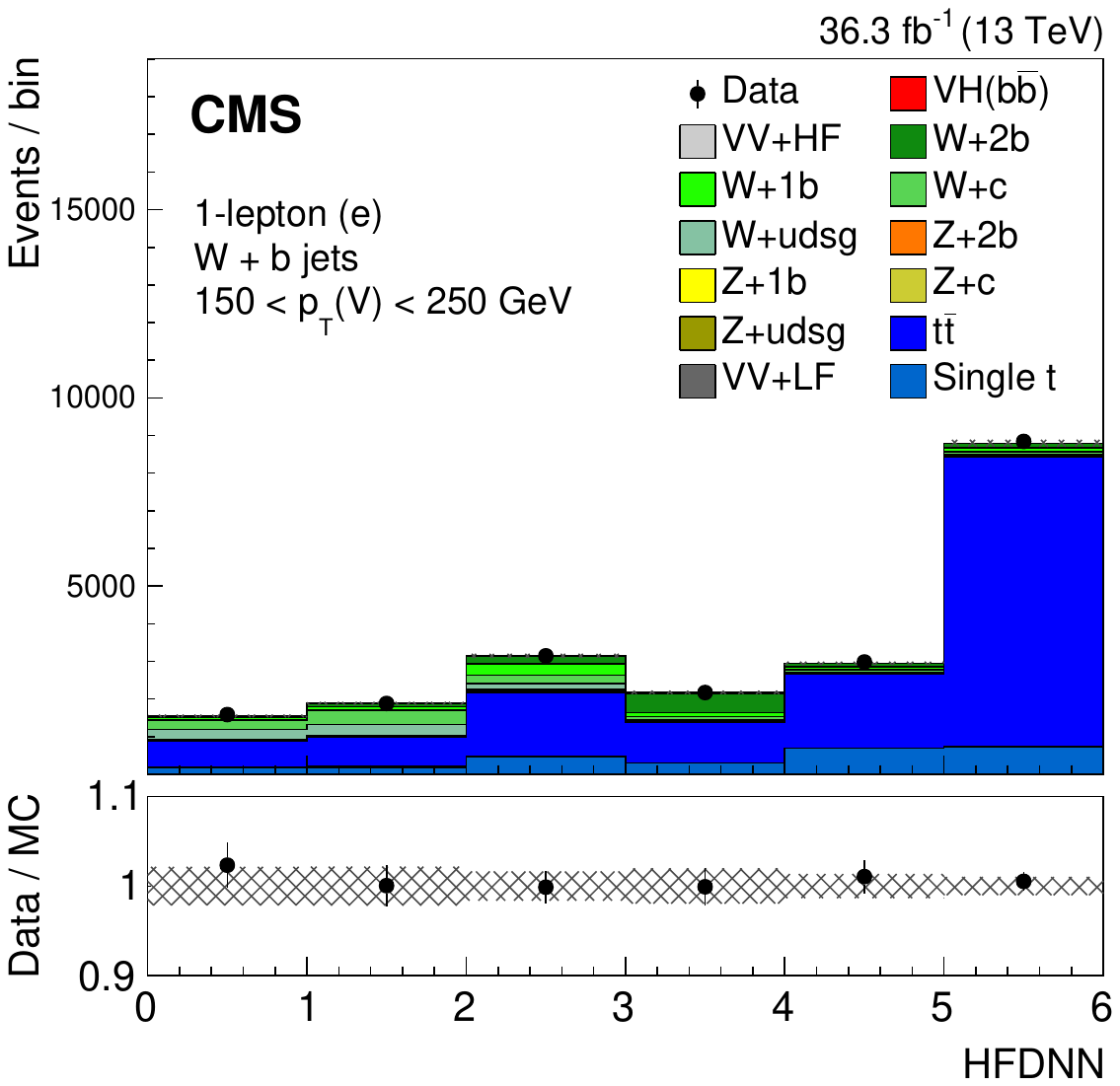}       
\caption{Distribution of the HFDNN scores in the $\zerol$ (\cmsLeft) and $\onel$ (\cmsRight) $\PZ$+$\PQb$ and $\PW$+$\PQb$ heavy-flavor CRs for the 2016 data set, after the fit to data. The output nodes target enrichment in the $\PV$+light-quark (first bin), $\PV$+$\PQc$ (second bin), $\PV$+$\PQb$ (third bin), $\PV$+$\PQb\PAQb$ (fourth bin), single top quark (fifth bin), and $\ttbar$ (sixth bin) backgrounds. The lower plots display the ratio of the data to the MC expectations.  The vertical bars on the points represent the statistical uncertainty in the ratio, and the hatched area shows the MC uncertainty.}
\label{fig:hfdnnplot}
\end{figure}

\begin{table*}[!htb]
\topcaption{Input variables used for the DNN training in the resolved SR of the 0-, 1-, and 2-lepton channels. Reconstructed jets associated with the Higgs boson candidate are classified as leading (labeled with suffix $j_{\mathrm{1}}$) and subleading (labeled with suffix $j_{\mathrm{2}}$) based on their \PQb tag score.}
\centering
\renewcommand*{\arraystretch}{1.2}
\cmsTable{
\begin{scotch}{llccc}
Variable & Description & $\zerol$ & $\onel$ & $\twol$ \\
 \hline
 $\Mjj$ & Dijet invariant mass & \checkmark	& \checkmark & \checkmark \\
 $\pt$(jj) & Dijet transverse momentum & \checkmark & \checkmark	& \checkmark \\
 $\ptvecmiss$  & Missing transverse momentum	&	\checkmark&	\checkmark & \checkmark  \\
 $M_{\text{t}}$(V)  & Transverse mass of the vector boson &	 &	\checkmark &     \\
 $\ptV$  & Transverse momentum of the vector boson &	 &	\checkmark & \checkmark   \\
 $\pt(\mathrm{jj})$/ $\ptV$  & Ratio of transverse momenta of the dijet system and the vector boson &	 &	\checkmark & \checkmark   \\
 $\dphiVjj$ & Azimuthal angle between the vector boson and the dijet directions	& \checkmark & \checkmark & \checkmark\\ 
 $\text{btag}(\mathrm{j_1})$& \PQb tagging score of leading jet &\checkmark&	\checkmark & \checkmark \\
 $\text{btag}(\mathrm{j_2})$& \PQb tagging score of subleading jet &\checkmark& \checkmark & \checkmark \\
 $\Delta\eta$(\jj) & Pseudorapidity difference between leading and subleading jet & \checkmark	&	\checkmark	&	\checkmark \\ 
 $\Delta\phi$(\jj) & Azimuthal angle between leading and subleading jet & \checkmark & \checkmark	&  \\ 
 \multirow{2}{*}{$\pt^{\text{max}}$($\mathrm{j_1,j_2}$)} & Maximum transverse momentum of jet &\multirow{2}{*}{\centering\checkmark} & \multirow{2}{*}{\centering\checkmark} &		\\
                                        & between leading and subleading jet & & &		\\
 SA5 & Number of soft-track jets with momentum greater than 5 \GeV	& \checkmark	& & \checkmark  \\
 $N_{\text{aj}}$ & Number of additional jets	& \checkmark	& \checkmark  &   \\
 $\text{btag}_{\text{max}}$ (add) & Maximum \PQb tagging discriminant score among additional jets	& \checkmark &  & \\ 
 $\pt^{\text{max}}$(add) & Maximum transverse momentum among additional jets & \checkmark &  & \\
 $\Delta\phi(\text{jet},\ptvecmiss)$ & Azimuthal angle between additional jet and $\ptvecmiss$ & \checkmark	 &  & \\
 $\Delta\phi(\text{lep},\ptvecmiss)$ & Azimuthal angle between lepton and $\ptvecmiss$ & 	 & \checkmark &   \\ 
 $M_{\PQt}$ & Reconstructed top quark mass & &\checkmark & \\ 
$\pt(\mathrm{j_1})$ & Transverse momentum of leading jet & & &\checkmark\\ 
 $\pt(\mathrm{j_2})$ & Transverse momentum of subleading jet & & &\checkmark \\
 $M(\PV)$ & Reconstructed vector boson mass & & &\checkmark \\ 
 $\Delta R (\PV,\jj)$ & Angular separation between the vector boson and the dijet system & & &\checkmark\\ 
 \multirow{2}{*}{$\Delta R (\PV,\jj)$ (kin)} & Angular separation between the vector boson and  & & &\multirow{2}{*}{\centering\checkmark}\\
  & the dijet system (reconstructed after kinematic fit) & & &\\
 $\sigma({M(\mathrm{jj})})$ & Resolution of dijet invariant mass & & &\checkmark\\
 $N_{\text{rec}}$ & Number of recoil jets & & &\checkmark\\
\end{scotch}
}
\label{tab:DNNinputs_list}
\end{table*}
Among the most discriminating variables for all channels are $\Mjj$, $\ptV$, the number of additional jets, and the angular separation between the two jets forming the Higgs boson candidate. To estimate the feature importance for the DNN training, the classifier is retrained sampling from the distribution of all possible feature combinations. The relative feature importance is defined using the approximate median significance metric~\cite{deepl}. All trainings are repeated 10 times to average out fluctuations from randomized initial weights and the stochastic gradient descent. The mean value of the estimated significance is used to compare with the baseline. In the $\twol$ channel, both regressed variables and variables evaluated after the kinematic fit are employed. The \PQb tagging status of the jets is exploited by using the \DeepCSV working point information. The trainings are performed in categories defined to target particular STXS bins for all channels, and the subsequent evaluation is performed for the same STXS bins. In the 0- and 1-lepton channels, a multiclass DNN (HFDNN) is trained in the heavy-flavor $\PZ$+$\PQb$ and $\PW$+$\PQb$ CRs, to separate the different $\PV$+jets components (vector boson production associated with light-quark, \PQc, and \PQb jets), single top quark, and $\ttbar$ backgrounds. The same input features and DNN architecture as for the signal/background classification are used (as shown in Table~\ref{tab:DNNinputs_list}). Instead of labels for the signal/background classification, the output of the DNN is an $m$-dimensional vector of probabilities for the $m$ background classes. Figure~\ref{fig:hfdnnplot} shows the HFDNN discriminants in the 0- and 1-lepton heavy-flavor CRs, after a maximum likelihood fit to the data. This is a simultaneous fit of all SRs and CRs in the analysis. 

The DNN score is used as a discriminating variable in each resolved SR, while different strategies are used in the resolved CRs, as discussed in Section~\ref{sec:results}. 

\subsection{Analysis of the large-radius jet topology}
\label{sec:boosted}
The boosted topology is included in the measurement in addition to the resolved analysis and targets the two STXS $\ptV$ bins above 250\GeV (i.e., 250--400\GeV and $>400\GeV$). In the boosted event categories a single AK8 jet is used to reconstruct the Higgs boson candidate, with the \DeepAKviii \bbbar tagging algorithm applied to the Higgs boson candidate decay products. This accounts for the kinematic properties of the event in the region where the two AK4 jets start to overlap.

An important feature of the analysis is the usage of the \bbbar tagging algorithm, both as a selection variable to define SR and CRs, and as a discriminating input feature of the MVA discriminant trained in the SR. The output of the discriminant node against light-quark jets of the \DeepAKviii algorithm (DeepAK8bbVsLight) is used in this measurement, since it is observed to provide the highest sensitivity for signal/background separation in the boosted analysis region. The $\PH\to\PQb\PAQb$ boosted signal output node is calibrated by performing data and simulation efficiency measurements, as described in Section~\ref{sec:rec}. The corresponding uncertainties are used as prior constraints in the fit. No external efficiency measurements are available for high-momentum light-quark, \PQc, and \PQb jets stemming from top quark decays. These components are dominant in the $\PV$+light-quark jets, $\PV$+$\PQb$ jets, and $\ttbar$ analysis CRs. Therefore, the normalizations of these processes are extracted in situ in the combined fit by including unconstrained parameters that scale the normalizations of these processes in the CRs, as discussed in Section~\ref{sec:results}. 

The SRs and CRs for the 0-, 1-, and 2-lepton channels are defined separately. In addition to the application of the baseline event selection defining the lepton channels discussed in Section~\ref{sec:resolved} for the resolved analysis, the common preselection valid for all channels in the large-radius jet topology requires $m_{\mathrm{SD}}>50\GeV$, $\ptH>250\GeV$, and $\ptV>250\GeV$. Only jets with $\abs{\eta}<2.5$ are considered. We define three CRs, enriched in the main background processes ($\PV$+$\PQb$ jets, $\PV$+light-quark jets, and $\ttbar$), and one SR for the extraction of the boosted $\PH\to\PQb\PAQb$ signal. To enrich the SR and CRs in $\ttbar$ and $\PV$+$\PQb$ jets processes, the \DeepAKviii discriminant must be larger than 0.8. For the light-quark jets region, the  complementary requirement ($<0.8$) is used. The value of 0.8 was optimized to retain most of the signal while minimizing the background contamination. In the 0- and 1-lepton channels, events in the $\ttbar$ CRs are selected if they have at least one AK4 \PQb-tagged jet, identified using the medium \DeepCSV working point. No additional leptons are allowed in SR and CRs. The dijet soft-drop mass must be in the range 90--150 \GeV for the SR, outside this range for the $\PV$+$\PQb$ jets CR, and above 50\GeV for the $\PV$+light-quark jets and $\ttbar$ CRs. Since all of the \PZ boson decay products are visible in the $\twol$ final state, $M(\PV)$ must be in the range 75--105\GeV for the SR, $\PZ$+$\PQb$ jets, and $\PZ$+light-quark jets CRs, and outside this range for the $\ttbar$ CR. Table~\ref{tab:boostedTable} summarizes the SR and CR selections in the boosted analysis.

\begin{table*}[!htb]
\centering
\topcaption{Selection criteria for the SR and CRs in the boosted topology for 0-, 1-, and 2-lepton channels. The DeepAK8bbVsLight designation represents the \DeepAKviii discriminant for the light-quark flavor discrimination node. The $\Mjj$ and $M(\PV)$ variables have units of \GeV.}
\begin{scotch}{lcccc}
Variable & SR & $\PV$+$\PQb$ jets & $\PV$+light-quark jets & $\ttbar$ \\
\hline
0-lepton & & & &\\
[\cmsTabSkip]
DeepAK8bbVsLight& $>$0.8 & $>$0.8 & $<$0.8 & $>$0.8 \\
$\Mjj$& $\in$[90--150]& $\in$[50--90] or $\in$[150--250]  & $>$50 & $>$50 \\
$N_{\text{al}}$ & $=$0 & $=$0 & $=$0 & $=$0 \\
$\Naj$& $=$0 & $=$0 & $=$0 & $>$0 \\
[\cmsTabSkip]
1-lepton & & & &\\
[\cmsTabSkip]
DeepAK8bbVsLight& $>$0.8 & $>$0.8 & $<$0.8 & $>$0.8 \\
$\Mjj$& $\in$[90--150]& $\in$[50--90] or $\in$[150--250] & $>$50 & $>$50 \\
$N_{\text{al}}$ & $=$0 & $=$0 & $=$0 & $=$0 \\
$\Naj$& $=$0 & $=$0 & $=$0 & $>$0 \\
[\cmsTabSkip]
2-lepton & & & &\\
[\cmsTabSkip]
DeepAK8bbVsLight& $>$0.8 & $>$0.8 & $<$0.8 & $>$0.8 \\
$\Mjj$& $\in$[90--150]& $\in$[50--90] or  $\in$[150--250]  & $\in$[90--150] & $>$50 \\
$M(\PV)$& $\in$[75--105]& $\in$[75--105]  & $\in$[75--105] &  $\notin$[75--105]   \\
\end{scotch}
\label{tab:boostedTable}
\end{table*}

\subsubsection{Multivariate discriminant in the SR for the boosted topology}
\label{sec:bdt_boosted}

Boosted decision trees (BDTs) were trained in the SRs for the 0-, 1-, and 2-lepton channels to separate the boosted Higgs boson decay signal from the sum of all background processes. Input variable optimizations, as well as overtraining checks, were performed. The input features of the BDT training take into account the kinematic properties of the event and include observables related to the AK8 jet candidate, such as its soft-drop mass and $\pt$. The invariant mass of the dijet system, as well as additional variables used for the DNN training in the resolved topology, are also added for events in the overlap region between the boosted and resolved analyses. For purely boosted events, these variables are set to default values in the BDT training and are not considered further. This training strategy is found to improve the BDT sensitivity by approximately 25\% as a result of the very good $\Mjj$ resolution. Additionally, the properties of the reconstructed vector boson recoiling against the Higgs boson candidate, and the \DeepAKviii output node against light-quark-jet discrimination, are used as inputs in the training. The inclusion of the \DeepAKviii discriminant in the BDT training leads to an improvement of around 30\% in the expected sensitivity in this region. Because the data-to-simulation corrections for the \DeepAKviii \PQb tagger are extracted for fixed operating points of the \DeepAKviii discriminant and do not correct the full shape of the output, binned working point values of the discriminant are used in the training instead of its full shape. The definition of the optimal working point used in the training is found from a scan of the expected sensitivity when varying the choice of the \DeepAKviii working point. For each configuration, a dedicated BDT is trained. The BDT score is used as the fitted variable in the boosted SR, while the \DeepAKviii output node is used in the boosted CRs. The $\PV$+$\PQb$ jets, $\PV$+light-quark jets and $\ttbar$ CRs make use of two bins in \PQb tagging score. The bin boundaries correspond to the \PQb tagging working points: [0.8, 0.97, 1]. The $\PV$+light-quark jets CR uses two bins in the complementary region, [0, 0.4, 0.8].

\section{Systematic uncertainties}
\label{sec:systematics}

Several systematic uncertainties affect the normalizations of the simulated signal and background processes, as well as the shapes of the multivariate discriminants fitted in the SRs. 

The theoretical uncertainty in the $\PH\to\PQb\PAQb$ branching fraction is 0.5\%~\cite{deFlorian:2016spz}. Theoretical uncertainties in the inclusive production cross sections are derived from factorization ($\mu_{\mathrm{F}}$) and renormalization ($\mu_{\mathrm{R}}$) scale variations and amount to 0.7, 0.6, and 25\% for the $\PW\PH$, quark-induced $\PZ\PH$, and gluon-induced $\PZ\PH$ processes, respectively~\cite{deFlorian:2016spz}. Migration uncertainties between the STXS bins are evaluated by studying the effect on the total cross section above each $\ptV$ and $\Naj$ boundary by varying $\mu_{\mathrm{F}}$ and $\mu_{\mathrm{R}}$ individually up and down by a factor of two around their default values and neglecting the largest variations. The largest change is used as the absolute uncertainty from this source. This change is then employed to increase the process normalization in STXS bins above the given $\ptV$ or $\Naj$ boundary, and decreasing it in the bins below. These uncertainties are in the range of 3--11 and 30--40\% for the quark- and gluon-induced processes, respectively, depending on the STXS bin considered. The gluon-induced process uncertainty is larger than for the quark-induced process because of the larger bin migration in the former case. Acceptance effects within each STXS bin are taken into account by varying $\mu_{\mathrm{F}}$ and $\mu_{\mathrm{R}}$, including a normalization factor to ensure these variations do not change the overall cross section of that particular STXS bin.

Theoretical uncertainties due to the choice of PDFs and the value of the strong coupling constant are derived for each signal and background process following the recommendations given in Ref.~\cite{pdfLHC} and are fully correlated across data-taking years. They are 1.9\% for the quark-induced $\PZ\PH$ and $\PW\PH$ processes, and 2.4\% for the gluon-induced $\PZ\PH$ process. Uncertainties coming from the variations of $\mu_{\mathrm{F}}$ and $\mu_{\mathrm{R}}$ are applied to all background processes and are fully correlated across data-taking years. The NLO EW corrections to the $\PZ\PH$ and $\PW\PH$ processes have a 2\% systematic uncertainty~\cite{deFlorian:2016spz}. The uncertainties in the diboson and single top quark production cross sections in the high-$\ptV$ region are both set to 15\%. These uncertainties are derived from CMS measurements of these processes~\cite{wzSM,zzSM}. However, they are increased to account for the different in the phase space between the present analysis and the previous measurement. 

\begin{figure}[!htb]  
\centering
\includegraphics[width=0.4\textwidth]{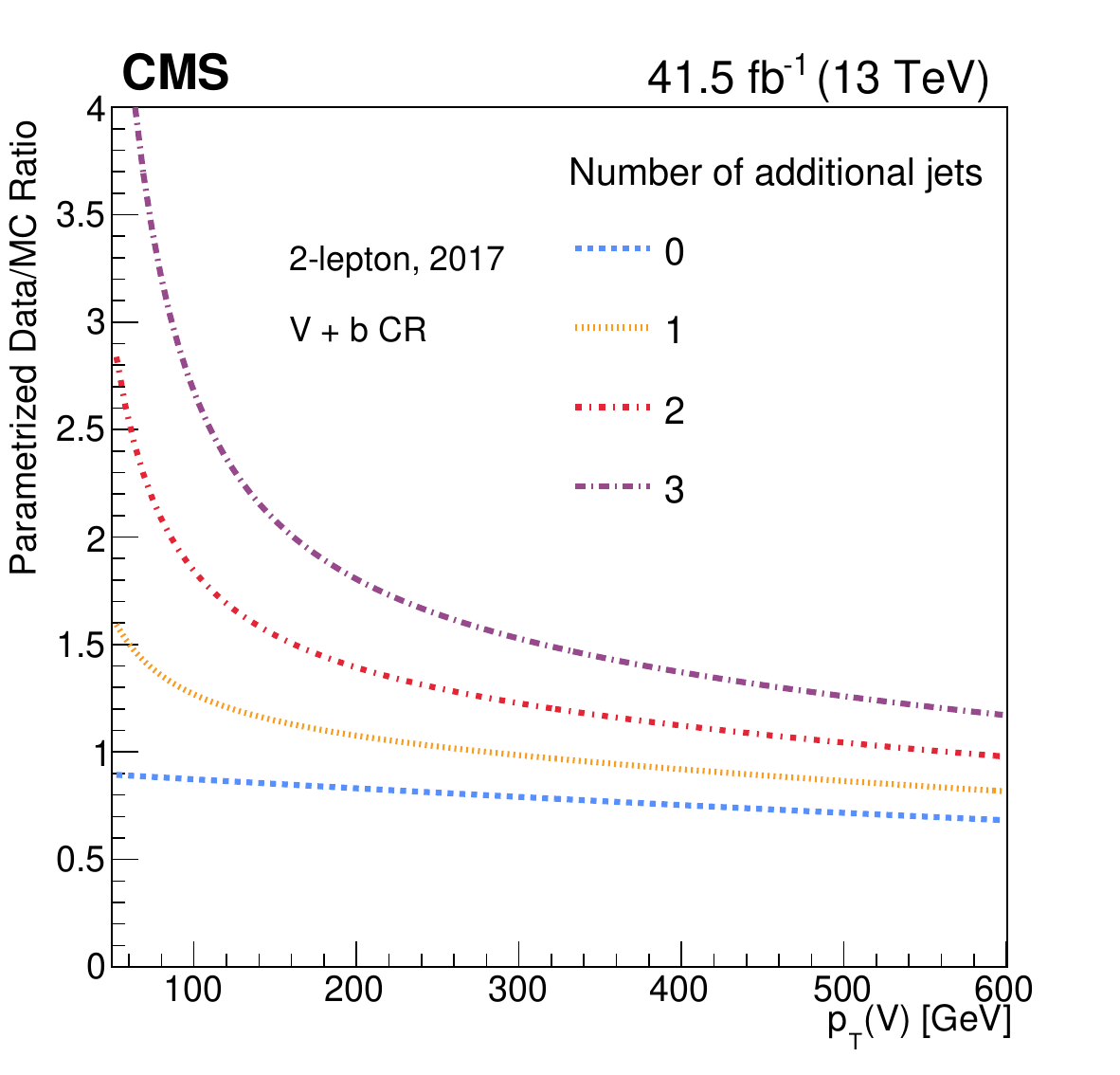}
\includegraphics[width=0.4\textwidth]{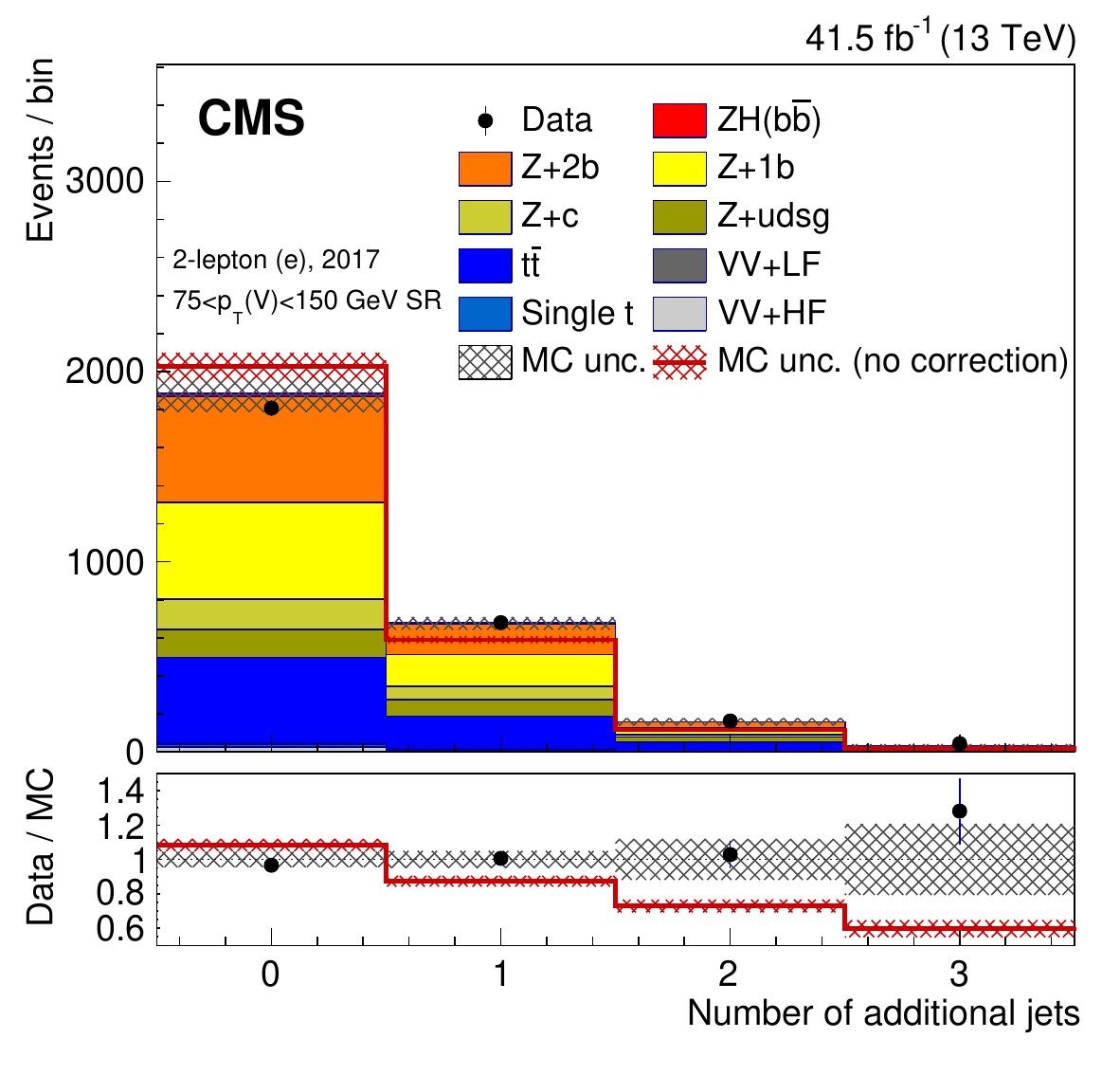}
\caption{The data-to-simulation ratio in the $\twol$ $\PZ$+$\PQb$ CR is parametrized as a two-dimensional function of the number of additional jets and $\ptV$. The values of the parametrization are shown as a function of these variables in the \cmsLeft panel. The distribution of the number of additional jets in the low-$\ptV$ STXS SR of the $\twol$ channel with (histogram) and without (red line) the application of the reweighting correction and associated systematic uncertainty is shown in the \cmsRight panel for the 2017 data set, after the likelihood fit to data. In the ratio pad, the hatched bands associated with the corrected (grey) and non-corrected (red) number of additional jets indicate the systematic uncertainty in the sum of the signal and background templates.}
\label{fig:rwNaddJet}
\end{figure}

The shapes of various distributions of the $\PV$+jet processes are derived by comparing simulation to data in control regions. The uncertainties affecting the $\PV$+jets processes impact both the process normalization and the shape of the distributions. The reweighting in the number of additional jets, discussed in Section~\ref{sec:rec}, is derived from the data-to-simulation ratio in the $\twol$ $\PZ$+$\PQb$ CR as a function of the number of additional jets and $\ptV$, which is shown in Fig.~\ref{fig:rwNaddJet} (\cmsLeft panel). The full value of the weight is used as the shape uncertainty associated with the reweighting. The uncertainties are treated as uncorrelated between data-taking years and across background processes to account for differences in the level of mismodeling between processes and data-taking periods. Figure~\ref{fig:rwNaddJet} (\cmsRight panel) shows an example distribution, after the likelihood fit to data, of the non-corrected and the corrected number of additional jets, for the 2017 data set in the low-$\ptV$ STXS region SR ($\twol$ channel).

The shape uncertainty associated with the number of additional jets reweighting correction, which is the full value of the weight, is included in the hatched grey band. Therefore, its size is significantly larger than the overall uncertainty associated with the non-corrected number of additional jets. The statistical uncertainty in the data yields is not included in the hatched bands and is present in the error bar associated with the data points in the ratio panel. The reweighting model, discussed in Section~\ref{sec:rec}, significantly improves the description of the data by the simulation in this observable. It also provides an additional uncertainty in this distribution.

Regarding the reweighting of the jet angular separation observable in the $\dRJJ < 1$ region discussed in Section~\ref{sec:sample}, shape variations are derived by extracting the statistical uncertainties of the $\dRJJ$ templates in data and simulation used for the reweighting procedure for each $\PV$+jets process. These shape uncertainties are treated as uncorrelated across analysis channels, STXS categories and data-taking periods. Normalization uncertainties in the NLO EW and NNLO QCD corrections applied to the high-$\ptV$ region in the signal and $\PV$+jets samples are estimated as 2 and 5\% for the NLO EW and NNLO QCD corrections, respectively.

The uncertainty in the integrated luminosity measurement is 1.2, 2.3, and 2.5\% in the 2016, 2017, and 2018 data-taking periods, respectively~\cite{CMS:2021xjt, CMS:2018elu, CMS:2019jhq}. These uncertainties are treated as partially correlated between the three data-taking years. A 4.6\% uncertainty in the total inelastic cross section~\cite{total_in_xs}, used to evaluate the PU profile in data for reweighting to the simulated PU profile, is applied.

The corrections applied to simulated samples to account for differences in the electron and muon trigger, reconstruction, and identification efficiencies with respect to data are affected by systematic uncertainties. These uncertainties originate from choices made in the efficiency measurement method, the selection applied to the leptons, and the limited size of the simulated samples used in the measurement. They depend on the lepton \pt and $\eta$, and affect the process normalizations by 1--2\%. An uncertainty in the $\ptmiss$ trigger efficiency correction is included in the model and has a $\approx$1\% effect. 

Uncertainties in the $\PQb$ tagging efficiency and misidentification rate measurements used in the analysis of the resolved topology depend on the jet flavor, and the \pt and $\eta$ of the jet. These uncertainties are split into 9 independent sources and 15 \pt and $\eta$ bins, and treated as uncorrelated between the three data-taking periods. For the $\bbbar$ tagging efficiencies used in the analysis of the boosted topology, uncertainties are provided for the $\PH\to\PQb\PAQb$ signal output node of the \DeepAKviii discriminant. They account for the limited size of the simulated samples available for the calibration; the relative uncertainty in the fraction of boosted jet contributions present in the calibration; and for the normalizations of these contributions.
The \bbbar tagging uncertainties are uncorrelated between the efficiency working points and are parametrized in regions of jet \pt (200--300, 300--400, 400--500, 500--600, $>$600)\GeV. 

Uncertainties in the JES and JER depend on the \pt and $\eta$ of the jets, and affect the kinematic properties of resolved and boosted jets, as well as the \ptmiss in the event. The uncertainties in the JES are split into independent sources~\cite{CMS:2016lmd} accounting for different experimental effects. Some of these uncertainty sources are correlated between the different data-taking periods, \eg, when that uncertainty component depends on the size of the available data sample. For \PQb-tagged jets, to which the previously described \PQb jet energy regression is applied, dedicated uncertainties in the JES and JER corrections are applied, as discussed in Section~\ref{sec:rec}.

Uncertainties affecting the shape of the $\ptV$ spectrum are taken into account. The discriminants used for the signal extraction in the resolved and boosted SRs are $\ptV$-dependent and $\ptV$ represents an important source of discrimination between signal and background, as discussed in Sections~\ref{sec:resolvednn}\ and~\ref{sec:boosted}. Additionally, the STXS categorization is based on $\ptV$, as described in Section~\ref{sec:resolved_stxscats}. Because the shape of the $\ptV$ spectrum is used in the fit model, the parameters that scale the background process normalizations are parametrized continuously in $\ptV$ rather than in individual STXS bins. To account for differences between adjacent categories, uncertainties in the \pt shapes are defined for each $\ptV$ boundary. These $\ptV$ uncertainties are parametrized with linear variation shape uncertainties are anticorrelated, following the STXS categorization at the $\ptV$ boundaries of 150\GeV (2-lepton channel only) and 250\GeV (all channels) for all processes. These uncertainties are also considered as uncorrelated across lepton channels, background processes, and data-taking eras. They are implemented with large prior constraints in the fit model to mimic freely floating flat prior uncertainties and are significantly constrained by the fit to data.

In addition to the $\ptV$-based uncertainties, freely floating parameters that scale the normalizations of the main background components, \ie, production of a vector boson associated with heavy (\PQb, \PQc) or light-quark jets and $\ttbar$, are included in the fit as described in Section~\ref{sec:results}. These parameters are constrained in the dedicated background-enriched regions discussed in Section~\ref{sec:eventselection}, \ie, $\PV$+$\PQb$ jets, $\PV$+light-quark jets, and  $\ttbar$. The parameters that scale the process normalizations are treated as uncorrelated between lepton flavors (\Pe, \PGm). 

To account for the finite sizes of the simulated samples, each bin of the simulated signal-plus-background template is allowed to vary within its statistical uncertainty, independently from the other bins in the distribution, following the Barlow--Beeston ``light" approach~\cite{bbb}. The impact of the systematic uncertainties in the measured cross sections in the different STXS bins is discussed in Section~\ref{sec:results}.

\section{Analysis results}
\label{sec:results}
The analysis targets the measurement of the Higgs boson signal strength using $\PV\PH$ production with a subsequent $\PH\to\PQb\PAQb$ decay and interprets the results in terms of the STXS categorization for the V(leptonic)H process, as discussed in Section~\ref{sec:resolved_stxscats}. Results are extracted from a simultaneous maximum likelihood fit of the signal-plus-background model to the data distributions in all SRs and CRs, based on the templates detailed in Table~\ref{tab:srcr_discrvar}.   

\begin{table*}[!htb]
\topcaption{Discriminating variables fitted in each SR and CR. The DeepAK8bbVsLight designation represents the \DeepAKviii discriminant for the light-quark flavor discrimination node.}
\centering
\cmsTable{
\begin{scotch}{ccccc}
& SR & $\ttbar$ CR & $\PV$+light-quark jets CR & $\PV$+$\PQb$ jets CR\\
\hline
0-lepton, resolved  & DNN & $\ptV$ & $\ptV$ & HFDNN\\
0-lepton, boosted & BDT & DeepAK8bbVsLight  & DeepAK8bbVsLight  & DeepAK8bbVsLight \\
1-lepton, resolved  & DNN & $\ptV$  & $\ptV$  & HFDNN\\
1-lepton, boosted  & BDT & DeepAK8bbVsLight & DeepAK8bbVsLight & DeepAK8bbVsLight \\
2-lepton, resolved  & DNN & $\ptV$  & $\ptV$  & DeepCSV scores\\
2-lepton, boosted  & BDT & DeepAK8bbVsLight & DeepAK8bbVsLight  & DeepAK8bbVsLight \\
\end{scotch}
}
\label{tab:srcr_discrvar}
\end{table*}

In the resolved and boosted SRs, the DNN and BDT classifiers presented in Section~\ref{sec:eventselection} are employed. The variable used in the resolved CRs depends on the channel. In the 0- and 1-lepton channels, the multiclassifier DNN described in Section~\ref{sec:resolved} is employed in the $\PV$+$\PQb$ jets CR, while $\ptV$ is used in the $\ttbar$ and $\PV$+light-quark jets CRs. In the $\twol$ channel, the fitted variable in the $\PV$+$\PQb$ jets CR is the score of the \DeepCSV discriminant, binned to align with the established working points, while the remaining CRs use $\ptV$ as in the 0- and 1-lepton channels. In the boosted CRs, the \DeepAKviii discriminant for the light-quark jet discrimination node (DeepAK8bbVsLight) is used in conjunction with the selection requirement described in Section~\ref{sec:boosted}. For the $\PV$+$\PQb$ jets and $\ttbar$ CRs, the \DeepAKviii discriminant is tagged as in the SR, while for the $\PV$+light-quark jets CR, the \DeepAKviii discriminant is required to be $<0.8$.

The analysis regions are partitioned into categories targeting specific STXS bins in order to maximize the sensitivity to the different STXS bin signals, as discussed in Section~\ref{sec:eventselection}. The shapes and normalizations of all distributions for the signal and background components are allowed to vary within the systematic uncertainties described in Section~\ref{sec:systematics}. These uncertainties are treated as independent nuisance parameters in the fit to the data. For the nuisance parameters with shape-altering effects, alternative templates that correspond to a variation of $\pm1$ standard deviation of the associated nuisance parameter are used. 

To reduce the effects from statistical fluctuations on these alternative templates in the SRs, a smoothing technique is applied to templates exhibiting the largest fluctuations with respect to the nominal templates. The normalization of the systematic variation is fixed, and the ratio of the template with respect to the nominal is smoothed. The uncertainty sources that show the largest fraction of bin-to-bin fluctuations are the JES, JER, and PU uncertainties. For those uncertainties, the smoothing procedure is applied to all processes in the analysis SRs. Freely floating parameters, termed process scale factors, accounting for the difference in normalization between simulation and data for the main background processes, namely $\ttbar$, $\PV$+udsg, $\PV$+$\PQc$, $\PV$+$\PQb$, and $\PV$+$\PQb\PAQb$, are constrained in the CRs and SRs. These process scale factors are correlated across lepton channels. In the 0- and 2-lepton channels, the $\PV$+$\PQb$ and $\PV$+$\PQb\PAQb$ components are split by employing freely floating parameters. In the $\onel$ channel, a freely floating parameter for the $\PV$+$\PQb\PAQb$ process is used in addition to a prior constraint that governs the ratio of $\PV$+$\PQb$ to $\PV$+$\PQb\PAQb$. This implementation is employed because the number of $\PV$+$\PQb$ events in the $\onel$ channel is limited due to the tight \PQb tagging requirement applied in the selection. 

To allow for shape uncertainties of the $\ptV$ spectrum predictions in the categories targeting different STXS bins, linear variations as a function of the reconstructed $\ptV$ are created, as discussed in Section~\ref{sec:systematics}. Process scale factors correcting for the normalization difference between simulation and data for the main background processes are employed, together with this $\ptV$-based uncertainty, to correct the shape of the $\ptV$ spectrum in simulation to match data. 

Additional unconstrained parameters, used to measure flavor tagging scale factors in situ in the boosted analysis regions, are employed to account for the (mis)tagging efficiency difference between data and simulation for high-momentum light-quark, $\PQc$, and $\PQb$ jets. The procedure is described in more detail in Section~\ref{sec:boosted}. These parameters are treated as fully correlated between channels, and are not correlated with the background process scale factors. As discussed in Section~\ref{sec:systematics}, the process scale factors and in situ flavor tagging scale factors are fully uncorrelated between lepton flavors.

\begin{figure*}[!htb]  
\centering
\includegraphics[width=0.65\textwidth]{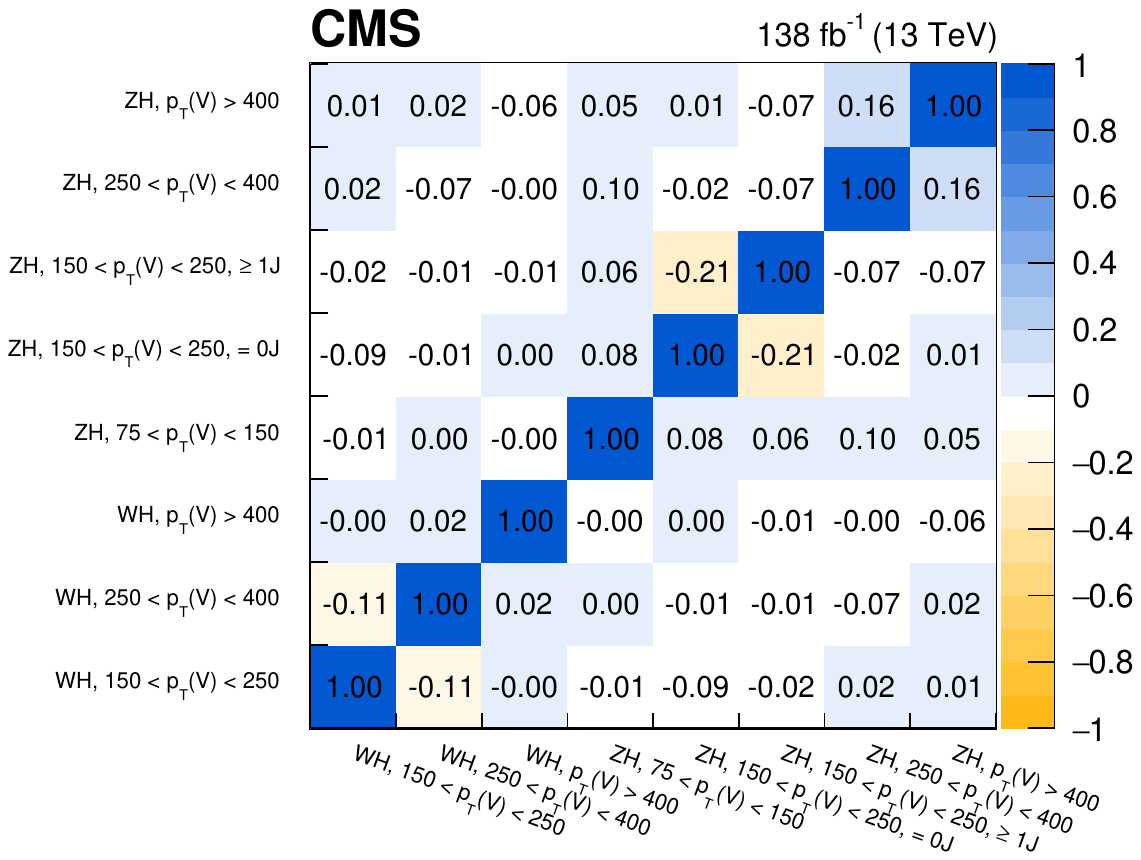} 
\caption{Correlation matrix of the parameters of interest in the STXS measurement. The vector boson momenta have units of \GeV.}
\label{fig:signalcomp}
\end{figure*}

\begin{figure*}[!htb]  
\centering
\includegraphics[width=0.8\textwidth]{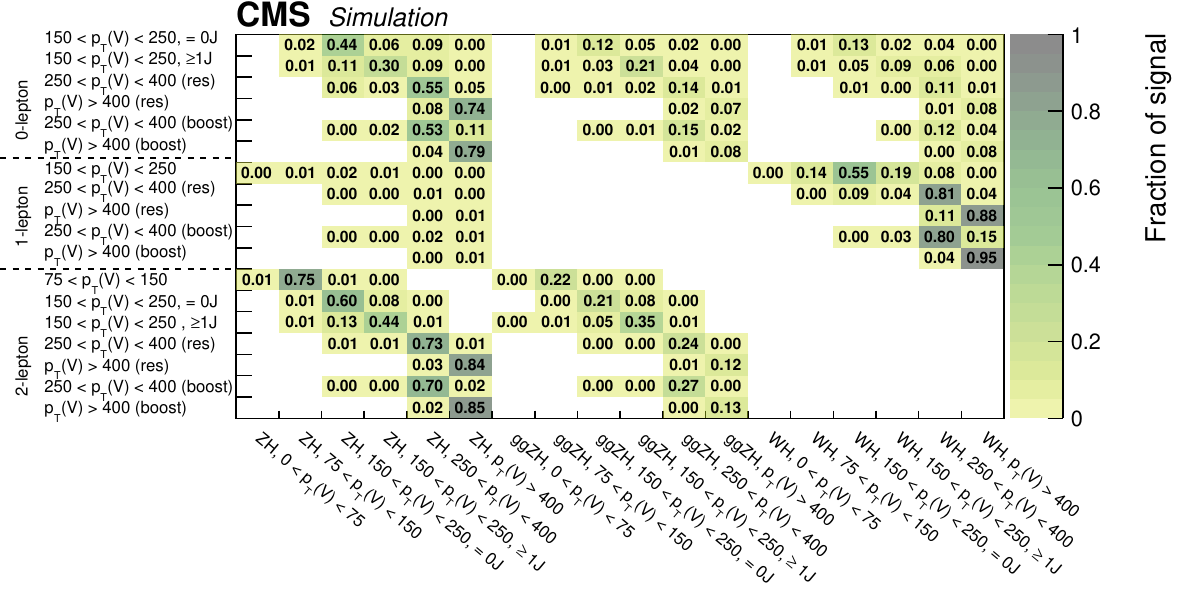} 
\caption{Contributions of the different STXS signal bins as a fraction of the total signal yield in each SR. The vector boson momenta have units of \GeV.}
\label{fig:signalcomp2}
\end{figure*}

Figure~\ref{fig:signalcomp} shows the correlation matrix of the signal strengths split by STXS bin for the analysis of all data-taking years combined. As expected (see Section~\ref{sec:resolved_stxscats}), the signal strengths for the medium $\ptV$ STXS bins with 0 and at least 1 jet exhibit the largest correlation ($-$21\%). The fractional contribution of each STXS bin to the total signal in each category is shown in Fig.~\ref{fig:signalcomp2}. The signal purity is higher in the $\twol$ channel than in the 0- and 1-lepton channels. 

\begin{figure}[!htb]  
\centering
\includegraphics[width=0.4\textwidth]{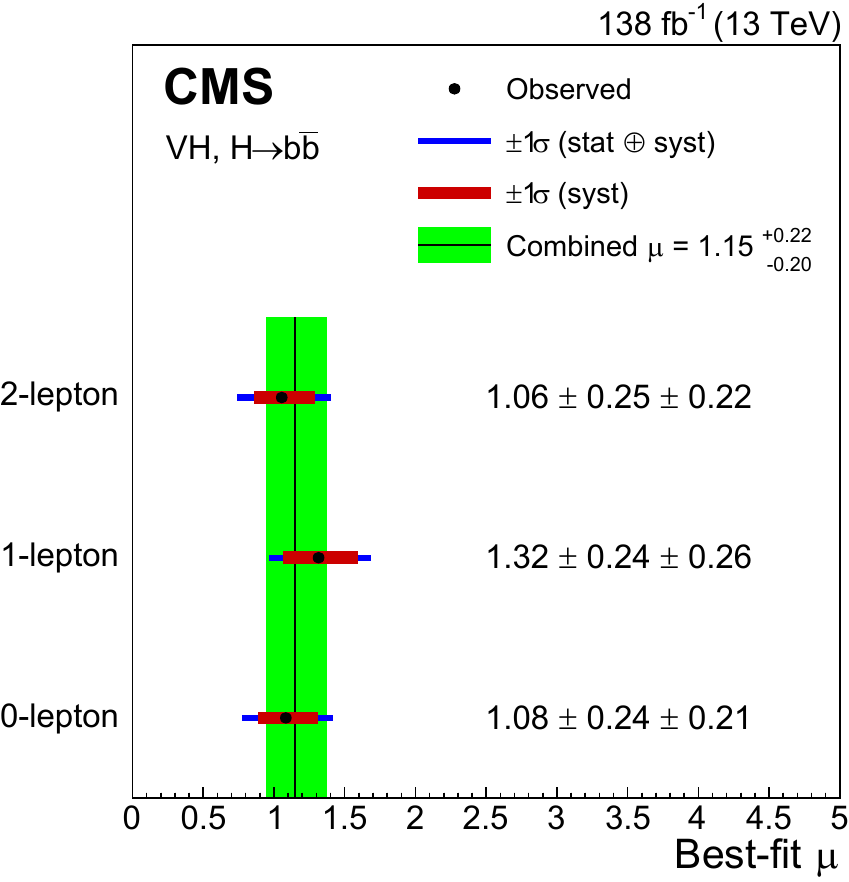}
\includegraphics[width=0.4\textwidth]{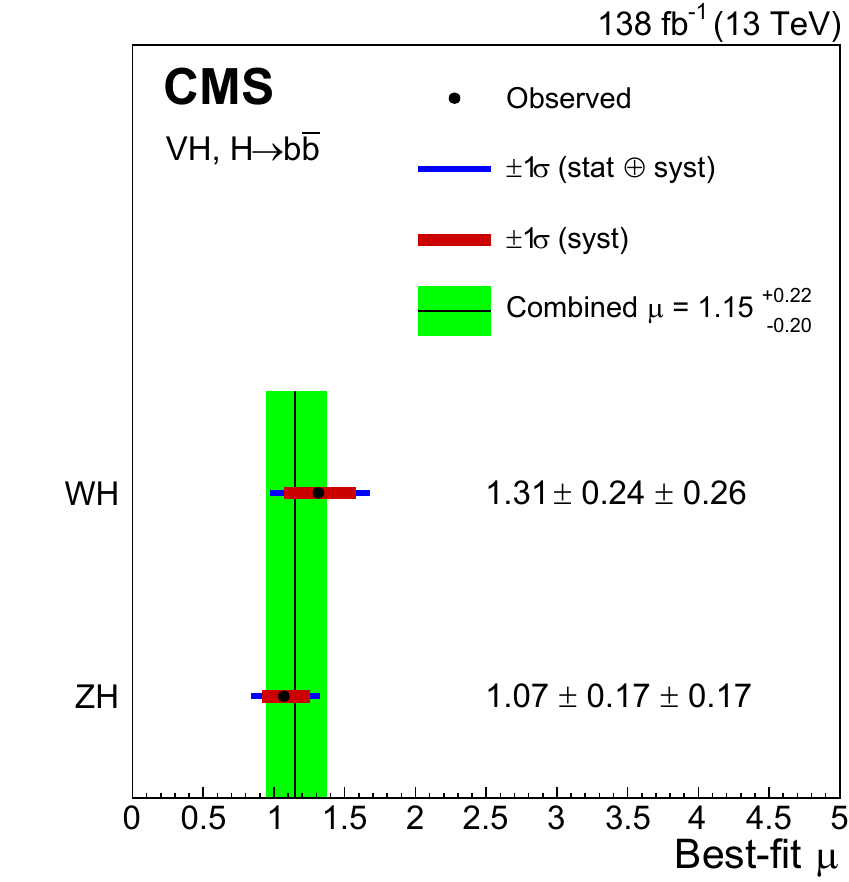}
\caption{Signal strengths (points) for the 0-, 1-, and 2-lepton channels (\cmsLeft) and the $\PZ\PH$ and $\PW\PH$ production modes (\cmsRight). The horizontal red and blue bars on the points represent the systematic and total uncertainties, respectively. The combined inclusive signal strength is shown by the vertical line, with the green band giving the 68\% confidence interval. The results combine the 2016--2018 data-taking years. The first and the second uncertainty values correspond to the statistical and systematic uncertainties, respectively.}
    \label{fig:perchan_plus_peryear}
\end{figure}

The inclusive signal strength extracted from a simultaneous maximum likelihood fit of the SRs and CRs, combining all three data-taking years, is $\mu=1.15^{+0.22}_{-0.20}$, where the uncertainties include both the statistical and systematic components. The individual signal strengths are $\mu=1.43 \pm 0.37$, $\mu=0.68 \pm 0.36$, and $\mu=1.23 \pm 0.30$ for the 2016, 2017, and 2018 data-taking years, respectively. Figure~\ref{fig:perchan_plus_peryear} shows the signal strengths per analysis channel, as well as the signal strengths split by production mode ($\PZ\PH$ or $\PW\PH$). The $p$-value compatibility of the individual deviations of the three analysis channels from the SM expectation ($\mu=1$) is 64\%, while the $p$-value compatibility of the three analysis channels with the inclusive $\PV\PH$, $\PH\to\PQb\PAQb$ signal strength is 84\%.  

\begin{figure*}[!htb]  
\centering
\includegraphics[width=0.6\textwidth]{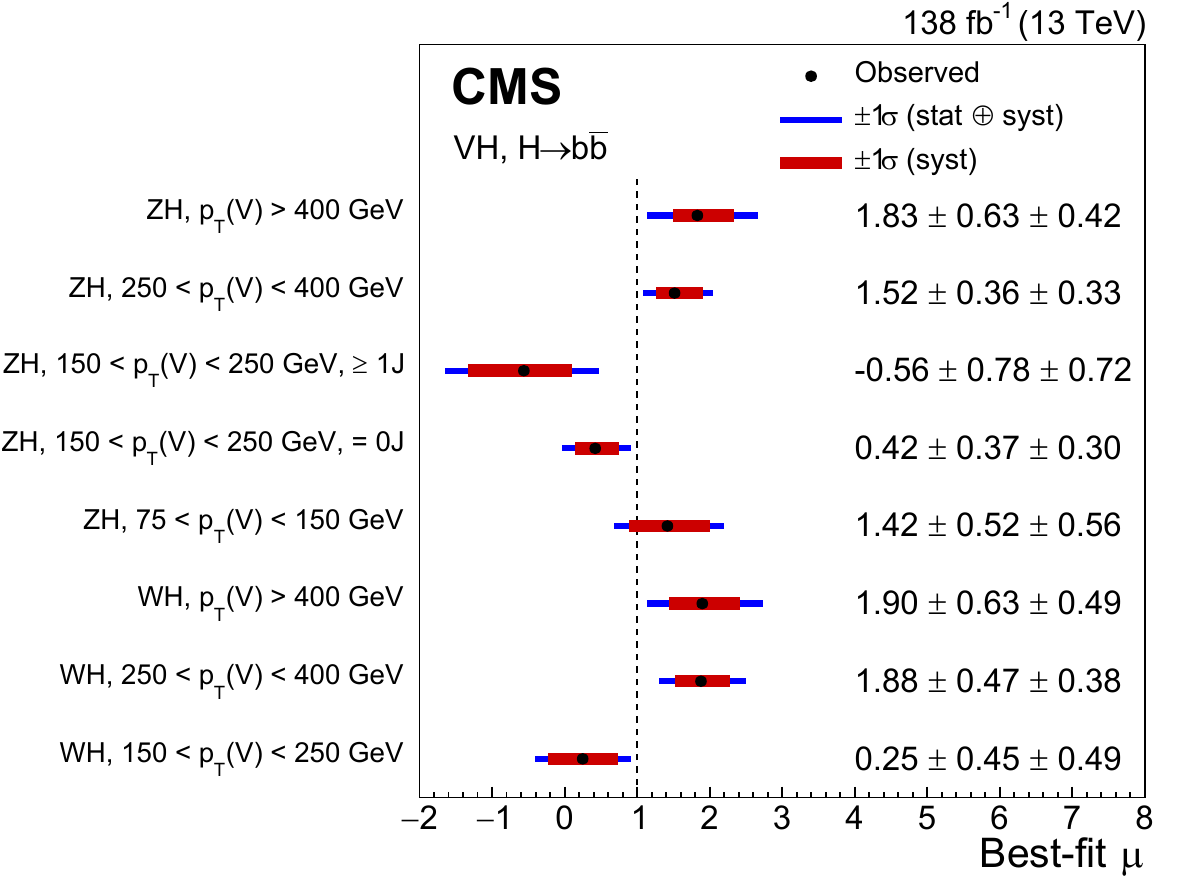} 
\caption{STXS signal strengths from the analysis of the 2016--2018 data. The vertical dashed line corresponds to the SM value of the signal strength. The first and the second uncertainty values correspond to the statistical and systematic uncertainties, respectively.}
 \label{fig:stxsresults}
\end{figure*}

\begin{figure*}[!htb]     
\centering
\includegraphics[width=0.6\textwidth]{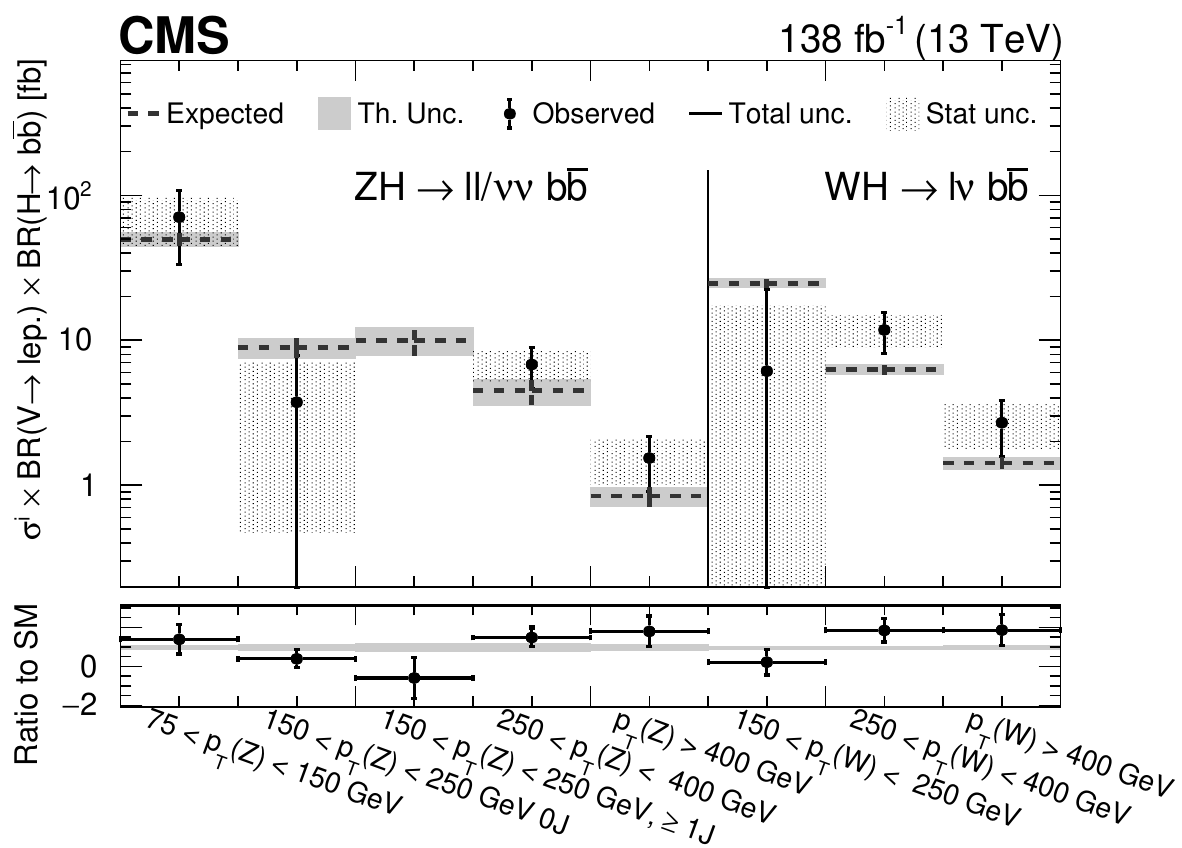} 
\caption{Measured values of $\sigma\mathcal{B}$, defined as the product of the $\PV\PH$ production cross sections multiplied by the branching fractions of \PV$\to$ leptons and $\PH\to\PQb\PAQb$, evaluated in the same STXS bins as for the signal strengths, combining all years. In the lower panel, the ratio of the observed results, with associated uncertainties, to the SM expectations is shown. If the observed signal strength for a given STXS bin is negative, no value is plotted for $\sigma\mathcal{B}$ in the upper panel.}
\label{fig:stxsresults2}
\end{figure*}

\begin{figure*}[!htb]  
\centering
\includegraphics[width=0.45\textwidth]{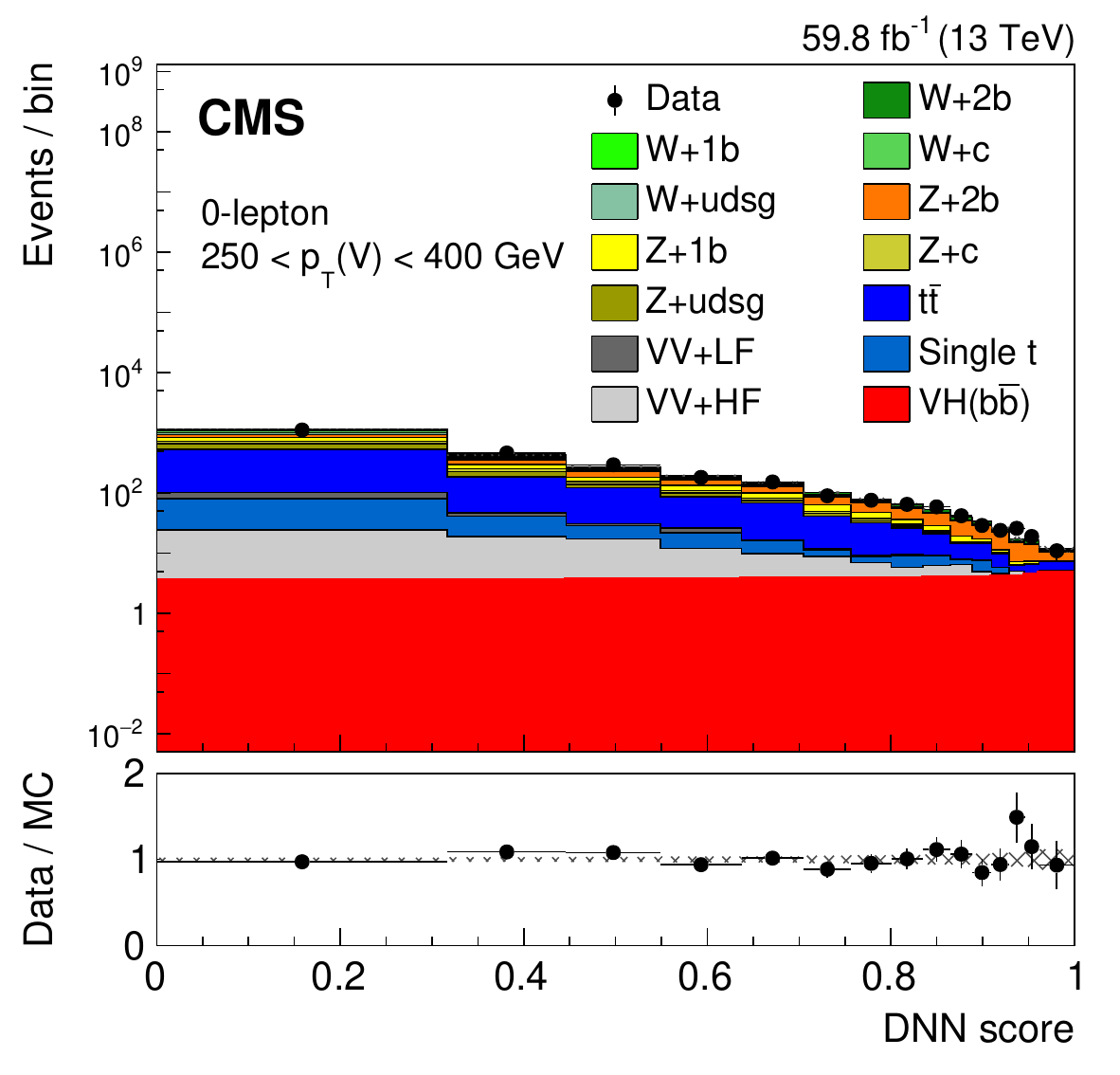} 
\includegraphics[width=0.45\textwidth]{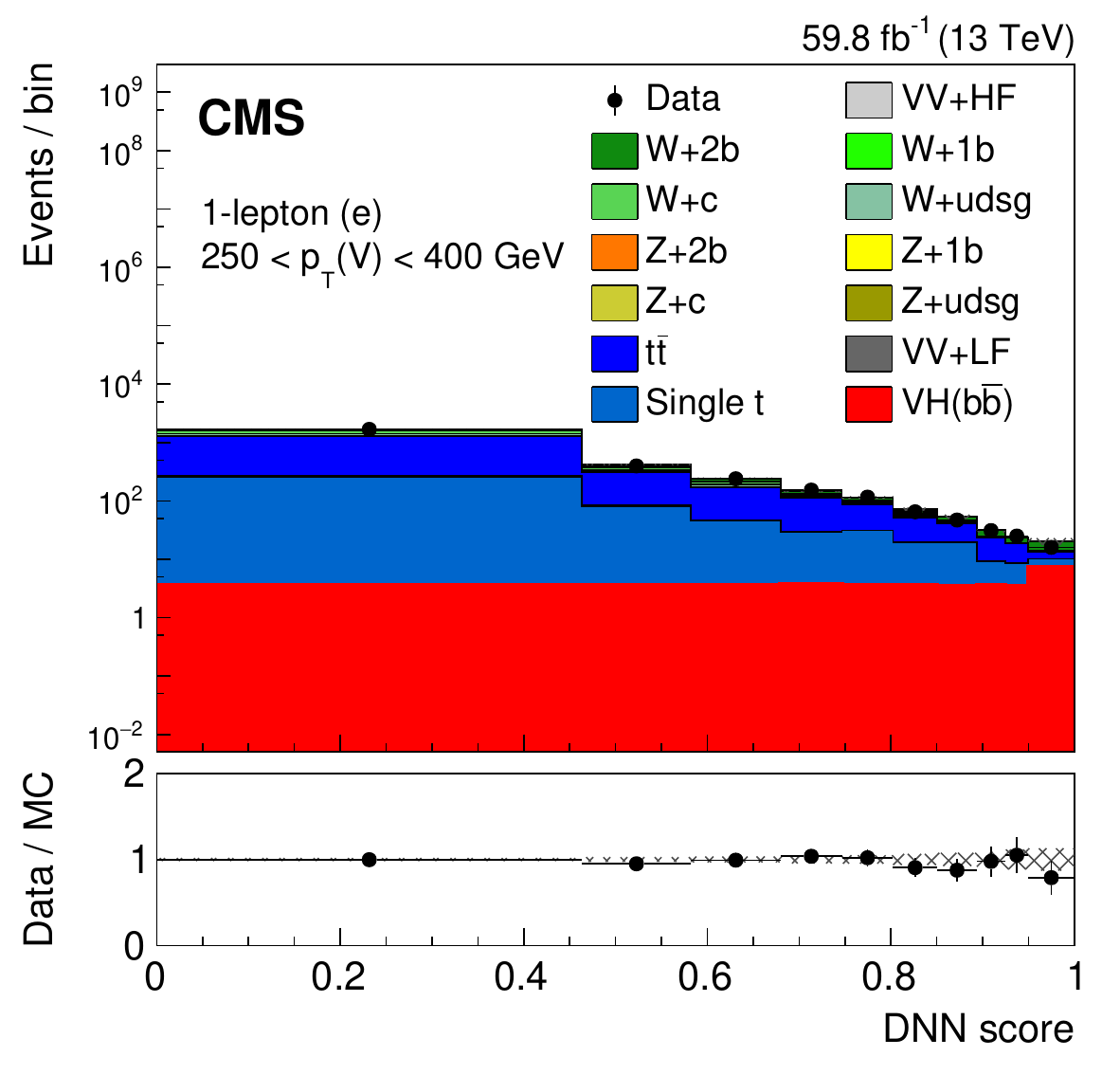} 
\includegraphics[width=0.45\textwidth]{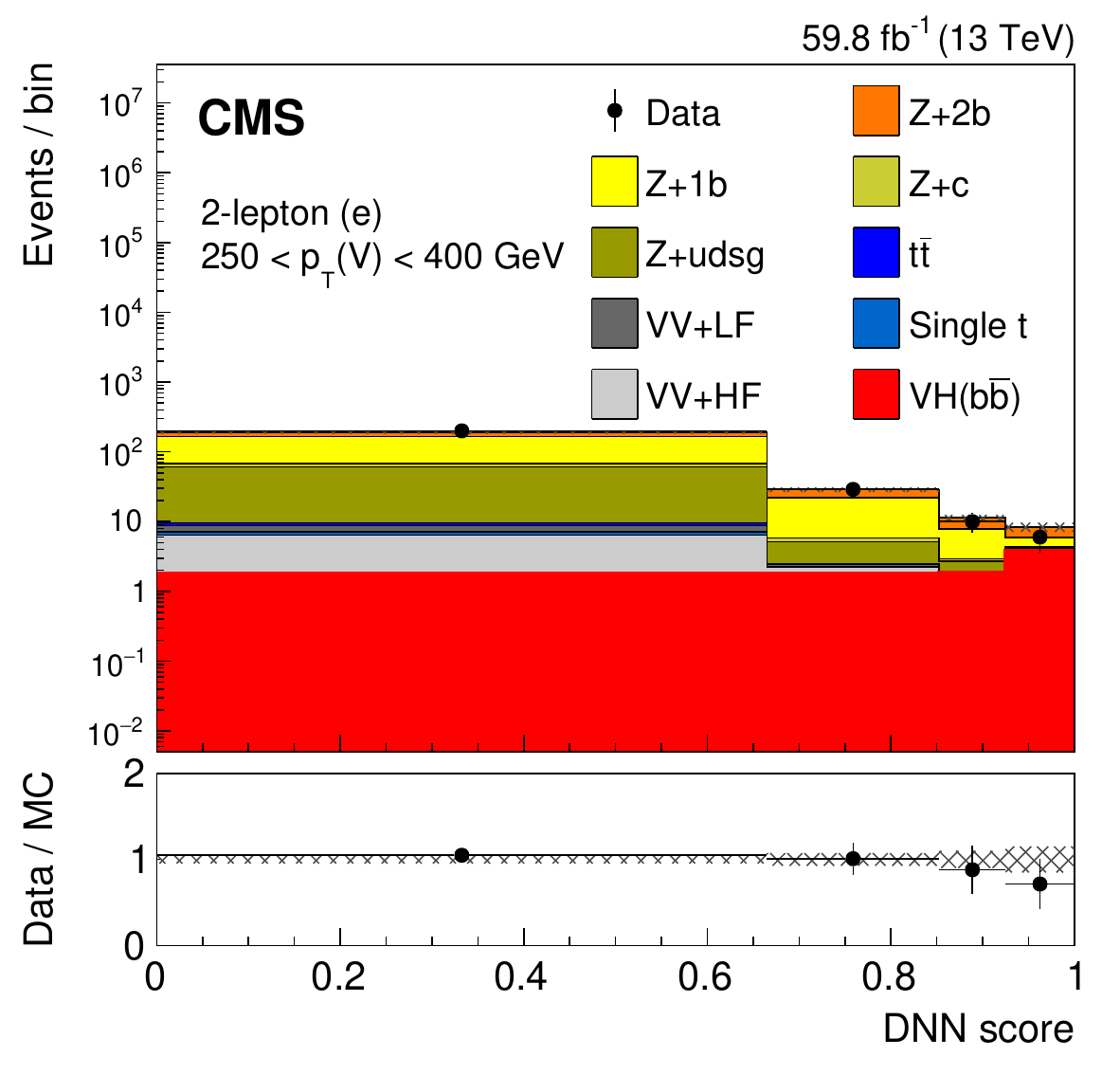}   
\caption{Post-fit distributions of the DNN discriminant in the $250<\ptV<400\GeV$ category of the $\zerol$ (top left), $\onel$ (top right) and $\twol$ (bottom) channels for the electron final state using the 2018 data set. The background contributions after the maximum likelihood fit are shown as filled histograms. The Higgs boson signal is also shown as a filled histogram, and is normalized to the signal strength shown in Fig.~\ref{fig:stxsresults}. The hatched band indicates the combined statistical and systematic uncertainty in the sum of the signal and background templates. The ratio of the data to the sum of the fitted signal and background is shown in the lower panel. The distributions that enter the maximum likelihood fit use the same binning as shown here.}
\label{fig:dnnplot}
\end{figure*}

\begin{figure}[!htb]  
\centering
\includegraphics[width=0.5\textwidth]{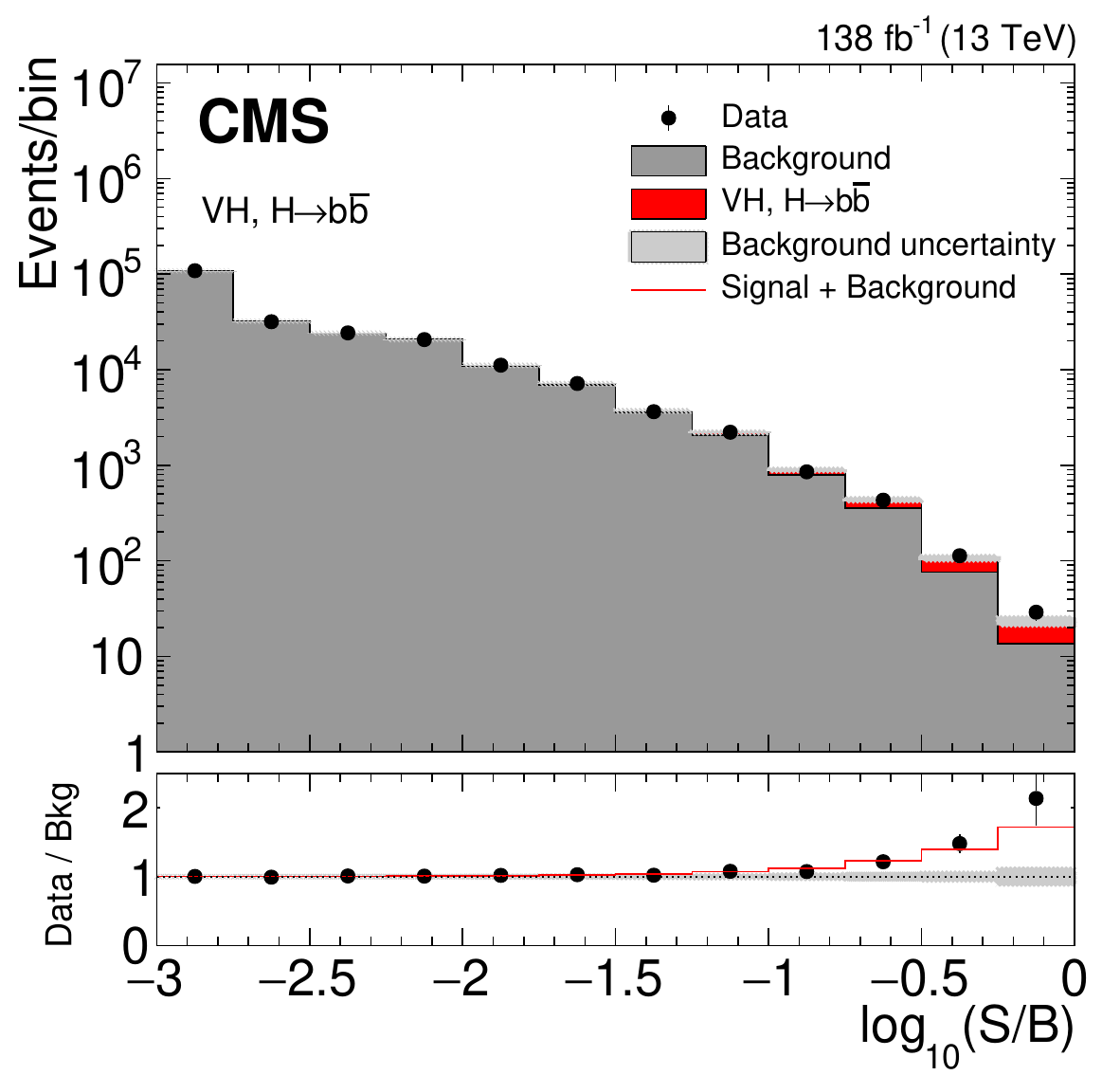} 
 \caption{Distributions of signal, background, and observed data event yields sorted into bins of similar signal-to-background ratio, as given by the result of the fit to the multivariate discriminants in the resolved and boosted categories. All events in the signal regions of the 2016--2018 data set are included. The red histogram indicates the Higgs boson signal assuming SM yields ($\mu=1$) and the sum of all backgrounds is given by the gray histogram. The lower panel shows the ratio of the observed data to the background expectation, with the total uncertainty in the background prediction indicated by the gray hatching. The red line indicates the sum of signal assuming the SM prediction plus background contribution, divided by the background.}
\label{fig:dnnsb}
\end{figure}

The measured signal strengths in the different STXS bins, fitting all data-taking years (2016--2018) are shown in Fig.~\ref{fig:stxsresults}. These results are interpreted in Fig.~\ref{fig:stxsresults2} as $\sigma\mathcal{B}$, the product of the production cross sections and the branching fractions for \PV$\to$ leptons and $\PH\to\PQb\PAQb$. To convert the results to measurements of the production cross section alone, theoretical uncertainties that modify the overall cross section of the individual STXS bins, or the inclusive cross section, are removed from the fit. These measured cross sections, along with the SM predictions, are given in Table~\ref{tab:stxs_xs}. The local inclusive observed (expected) significance of the measured $\PZ\PH$ and $\PW\PH$ signals, over the background-only expectation, is found to be 6.3 (5.6) standard deviations when taking into account all three data-taking years. Examples of post-fit distributions of the DNN output scores in the SRs of the 2018 data set are shown in Fig.~\ref{fig:dnnplot} for the 0-, 1-, and 2-lepton channels in the category targeting the $250<\ptV<400\GeV$ STXS bin. Figure~\ref{fig:dnnsb} shows the distribution of events in all channels, sorted according to the observed value of $\log_{10}$ (S/B), for the three data-taking years combined; here, the signal (S) and background (B) yields are determined from the discriminant scores used in the resolved and boosted analyses. 
 
\begin{table*}[!htb]
\centering
\topcaption{Predicted and measured values of the product of the cross section and branching fractions in the V(leptonic)H STXS process scheme. The SM predictions for each bin are calculated using the inclusive values reported in Ref.~\cite{deFlorian:2016spz}. The uncertainties shown are the combined statistical and systematic components.}
\begin{scotch}{lccc}
    STXS bin & Expected $\sigma\mathcal{B}$ [fb] & Observed $\sigma\mathcal{B}$ [fb] & Best-fit $\mu$ \\
    \hline
    ZH $75<\pt(\PZ)<150$ \GeV  & 50.0 $\pm$ 5.3 & 71 $\pm$ 38 & 1.4 $\pm$ 0.8  \\
    ZH $150<\pt(\PZ)<250$ \GeV 0 jets  & 9.0 $\pm$ 1.4 & 3.8 $\pm$ 4.1 & 0.4 $\pm$ 0.5  \\
    ZH $150<\pt(\PZ)<250$ \GeV $\geq$1 jets & 10.1 $\pm$ 2.2 & $<$0 & $-$0.6 $\pm$ 1.0  \\
    ZH $250<\pt(\PZ)<400$ \GeV  & 4.5 $\pm$ 0.9 & 6.9 $\pm$ 2.2 & 1.5 $\pm$ 0.5  \\
    ZH $\pt(\PZ)>400$ \GeV   & 0.9 $\pm$ 0.1 & 1.6 $\pm$ 0.6 & 1.8 $\pm$ 0.8  \\
    WH $150<\pt(\PW)<250$ \GeV & 24.9 $\pm$ 1.8 &  6 $\pm$ 16 & 0.2 $\pm$ 0.7 \\
    WH $250<\pt(\PW)<400$ \GeV  & 6.3 $\pm$ 0.5 & 11.9 $\pm$ 3.8 & 1.9 $\pm$ 0.6  \\
    WH $\pt(\PW)>400$ \GeV   & 1.4 $\pm$ 0.1 &  2.7 $\pm$ 1.1 &  1.9 $\pm$ 0.8  \\
    \end{scotch}
  \label{tab:stxs_xs}
\end{table*}

Table~\ref{tab:sys_comb} shows the contribution, in terms of absolute uncertainties, to the uncertainty in the measured inclusive signal strength originating from the various sources of systematic uncertainty. This contribution for a given group of uncertainties is defined as the difference in quadrature between the total uncertainty in the signal strength and the uncertainty in the signal strength with the nuisance parameters of the corresponding group fixed to their best-fit values. The total statistical uncertainty is defined as the uncertainty in the signal strength when all the constrained nuisance parameters are fixed to their best-fit values, while the total systematic uncertainty is defined as the difference in quadrature between the total uncertainty in the signal strength and the total statistical uncertainty. Table~\ref{tab:sys_comb} breaks the total uncertainty down into the following sources.

\begin{itemize}
\item Theoretical uncertainties in the signal and background components.
\item Limited size of simulated samples.
\item Simulation modeling, including uncertainty sources associated with the modeling of the $\PV$+jets background components, as discussed in Section~\ref{sec:sample}. Additionally, the $\ptV$ migration uncertainties are included in this category.
\item Experimental uncertainties (\PQb tagging, integrated luminosity, JES and JER, lepton identification, and trigger). The JES and JER components include the dedicated uncertainty in mass scale and smearing that is applied for jets subject to the \PQb jet energy regression.
\end{itemize}

The limited size of the NLO $\PV$+jets samples is the largest contribution to the overall $\PV\PH$, ${\PH\to\PQb\PAQb}$ signal strength uncertainty. 

\begin{table}[!htb]
\centering
\topcaption{The sources of systematic uncertainty in the inclusive signal strength measurement and their positive and negative values.}
\begin{scotch}{lc}
    &  $\Delta\mu$ \\
    \hline
    Background (theory)  & +0.043  $-$0.043\\
    Signal (theory)  & +0.088  $-$0.059\\
    [\cmsTabSkip]
    MC sample size  & +0.078  $-$0.078\\
    Simulation modeling  & +0.059  $-$0.059\\
    \PQb tagging  & +0.050  $-$0.046\\
    Jet energy resolution  & +0.036  $-$0.028\\
    Int. luminosity  & +0.032  $-$0.027\\
    Jet energy scale  & +0.025  $-$0.025\\
    Lepton ident. & +0.008  $-$0.007\\
    Trigger ($\ptvecmiss$)  & +0.002  $-$0.001\\
  \end{scotch}
  \label{tab:sys_comb}
\end{table}

\subsection{Cross-check analysis: extraction of  \texorpdfstring{$\PV\PZ$ with $\PZ\to\bbbar$}{VZ with Z to bbbar}}

The $\PV\PZ$ process, where the \PZ boson decays into a $\PQb\PAQb$ pair, has an identical final state to the $\PV\PH$ process with $\PH\to\PQb\PAQb$. Therefore, it is used to cross-check the methodology for the $\PV\PH$, $\PH\to\PQb\PAQb$ analysis. The DNN and BDT discriminants in the resolved and boosted SRs are trained using the simulated diboson sample as signal. All other processes are considered as background. The $\PV\PZ$, $\PZ\to\PQb\PAQb$ analysis makes use of the same event categorization as the $\PV\PH$ analysis discussed in Section~\ref{sec:eventselection}, with the only modification being the requirement that $\Mjj$ lies in the range 60--120\GeV to define the SR for all channels. The inclusive observed $\PV\PZ$, $\PZ\to\PQb\PAQb$ signal strength is $\mu=1.25\pm0.14$, corresponding to observed and expected significances well above 5 standard deviations. The per-production process signal strengths, $\mu_{\PZ\PZ}$ and $\mu_{\PW\PZ}$, are $1.19 \pm 0.09\stat \pm 0.11\syst$ and $1.61\pm 0.18\stat \pm 0.24\syst$, respectively.

\section{Summary}
\label{sec:summa}
Measurements are presented of the cross section for the associated production of the 125\GeV Higgs boson and a \PW or \PZ boson, where the Higgs boson decays to \bbbar and the vector bosons decay to leptons. Proton-proton collision data collected by the CMS experiment during 2016--2018 at ${\sqrt{s}=13\TeV}$ are used, corresponding to an integrated luminosity of 138\fbinv. Five decay channels are analyzed, and both resolved as well as merged-jet topology are employed in each vector boson decay mode. An additional subcategorization in the transverse momentum of the vector boson and the number of additional jets in the event is applied to maximize the sensitivity of different simplified template cross section bins. The overall signal strength, combining all analysis categories, is found to be $\mu=1.15^{+0.22}_{-0.20}$. The production of the Higgs boson in association with a vector boson and decays to bottom quark pairs is established with an observed (expected) significance of 6.3 (5.6) standard deviations. 

\begin{acknowledgments}
  We congratulate our colleagues in the CERN accelerator departments for the excellent performance of the LHC and thank the technical and administrative staffs at CERN and at other CMS institutes for their contributions to the success of the CMS effort. In addition, we gratefully acknowledge the computing centers and personnel of the Worldwide LHC Computing Grid and other centers for delivering so effectively the computing infrastructure essential to our analyses. Finally, we acknowledge the enduring support for the construction and operation of the LHC, the CMS detector, and the supporting computing infrastructure provided by the following funding agencies: SC (Armenia), BMBWF and FWF (Austria); FNRS and FWO (Belgium); CNPq, CAPES, FAPERJ, FAPERGS, and FAPESP (Brazil); MES and BNSF (Bulgaria); CERN; CAS, MoST, and NSFC (China); MINCIENCIAS (Colombia); MSES and CSF (Croatia); RIF (Cyprus); SENESCYT (Ecuador); MoER, ERC PUT and ERDF (Estonia); Academy of Finland, MEC, and HIP (Finland); CEA and CNRS/IN2P3 (France); SRNSF (Georgia); BMBF, DFG, and HGF (Germany); GSRI (Greece); NKFIH (Hungary); DAE and DST (India); IPM (Iran); SFI (Ireland); INFN (Italy); MSIP and NRF (Republic of Korea); MES (Latvia); LAS (Lithuania); MOE and UM (Malaysia); BUAP, CINVESTAV, CONACYT, LNS, SEP, and UASLP-FAI (Mexico); MOS (Montenegro); MBIE (New Zealand); PAEC (Pakistan); MES and NSC (Poland); FCT (Portugal); MESTD (Serbia); MCIN/AEI and PCTI (Spain); MOSTR (Sri Lanka); Swiss Funding Agencies (Switzerland); MST (Taipei); MHESI and NSTDA (Thailand); TUBITAK and TENMAK (Turkey); NASU (Ukraine); STFC (United Kingdom); DOE and NSF (USA).
  
  \hyphenation{Rachada-pisek} Individuals have received support from the Marie-Curie program and the European Research Council and Horizon 2020 Grant, contract Nos.\ 675440, 724704, 752730, 758316, 765710, 824093, and COST Action CA16108 (European Union); the Leventis Foundation; the Alfred P.\ Sloan Foundation; the Alexander von Humboldt Foundation; the Science Committee, project no. 22rl-037 (Armenia); the Belgian Federal Science Policy Office; the Fonds pour la Formation \`a la Recherche dans l'Industrie et dans l'Agriculture (FRIA-Belgium); the Agentschap voor Innovatie door Wetenschap en Technologie (IWT-Belgium); the F.R.S.-FNRS and FWO (Belgium) under the ``Excellence of Science -- EOS" -- be.h project n.\ 30820817; the Beijing Municipal Science \& Technology Commission, No. Z191100007219010 and Fundamental Research Funds for the Central Universities (China); the Ministry of Education, Youth and Sports (MEYS) of the Czech Republic; the Shota Rustaveli National Science Foundation, grant FR-22-985 (Georgia); the Deutsche Forschungsgemeinschaft (DFG), under Germany's Excellence Strategy -- EXC 2121 ``Quantum Universe" -- 390833306, and under project number 400140256 - GRK2497; the Hellenic Foundation for Research and Innovation (HFRI), Project Number 2288 (Greece); the Hungarian Academy of Sciences, the New National Excellence Program - \'UNKP, the NKFIH research grants K 124845, K 124850, K 128713, K 128786, K 129058, K 131991, K 133046, K 138136, K 143460, K 143477, 2020-2.2.1-ED-2021-00181, and TKP2021-NKTA-64 (Hungary); the Council of Science and Industrial Research, India; ICSC -- National Research Center for High Performance Computing, Big Data and Quantum Computing, funded by the NetxGenerationEU program (Italy); the Latvian Council of Science; the Ministry of Education and Science, project no. 2022/WK/14, and the National Science Center, contracts Opus 2021/41/B/ST2/01369 and 2021/43/B/ST2/01552 (Poland); the Funda\c{c}\~ao para a Ci\^encia e a Tecnologia, grant CEECIND/01334/2018 (Portugal); the National Priorities Research Program by Qatar National Research Fund; MCIN/AEI/10.13039/501100011033, ERDF ``a way of making Europe", and the Programa Estatal de Fomento de la Investigaci{\'o}n Cient{\'i}fica y T{\'e}cnica de Excelencia Mar\'{\i}a de Maeztu, grant MDM-2017-0765 and Programa Severo Ochoa del Principado de Asturias (Spain); the Chulalongkorn Academic into Its 2nd Century Project Advancement Project, and the National Science, Research and Innovation Fund via the Program Management Unit for Human Resources \& Institutional Development, Research and Innovation, grant B37G660013 (Thailand); the Kavli Foundation; the Nvidia Corporation; the SuperMicro Corporation; the Welch Foundation, contract C-1845; and the Weston Havens Foundation (USA).
\end{acknowledgments}
\bibliography{auto_generated}  
\cleardoublepage \appendix\section{The CMS Collaboration \label{app:collab}}\begin{sloppypar}\hyphenpenalty=5000\widowpenalty=500\clubpenalty=5000\input{HIG-20-001-public-authorlist.tex}\end{sloppypar}
\end{document}

%% file: HIG-20-001-public-authorlist.tex
\cmsinstitute{Yerevan Physics Institute, Yerevan, Armenia}
{\tolerance=6000
A.~Tumasyan\cmsAuthorMark{1}\cmsorcid{0009-0000-0684-6742}
\par}
\cmsinstitute{Institut f\"{u}r Hochenergiephysik, Vienna, Austria}
{\tolerance=6000
W.~Adam\cmsorcid{0000-0001-9099-4341}, J.W.~Andrejkovic, T.~Bergauer\cmsorcid{0000-0002-5786-0293}, S.~Chatterjee\cmsorcid{0000-0003-2660-0349}, K.~Damanakis\cmsorcid{0000-0001-5389-2872}, M.~Dragicevic\cmsorcid{0000-0003-1967-6783}, A.~Escalante~Del~Valle\cmsorcid{0000-0002-9702-6359}, P.S.~Hussain\cmsorcid{0000-0002-4825-5278}, M.~Jeitler\cmsAuthorMark{2}\cmsorcid{0000-0002-5141-9560}, N.~Krammer\cmsorcid{0000-0002-0548-0985}, L.~Lechner\cmsorcid{0000-0002-3065-1141}, D.~Liko\cmsorcid{0000-0002-3380-473X}, I.~Mikulec\cmsorcid{0000-0003-0385-2746}, P.~Paulitsch, J.~Schieck\cmsAuthorMark{2}\cmsorcid{0000-0002-1058-8093}, R.~Sch\"{o}fbeck\cmsorcid{0000-0002-2332-8784}, D.~Schwarz\cmsorcid{0000-0002-3821-7331}, M.~Sonawane\cmsorcid{0000-0003-0510-7010}, S.~Templ\cmsorcid{0000-0003-3137-5692}, W.~Waltenberger\cmsorcid{0000-0002-6215-7228}, C.-E.~Wulz\cmsAuthorMark{2}\cmsorcid{0000-0001-9226-5812}
% ...
\par}
\vskip\cmsinstskip
\dag:~Deceased\\
$^{1}$Also at Yerevan State University, Yerevan, Armenia\\
$^{2}$Also at TU Wien, Vienna, Austria\\
% ...
